\documentclass{natureprintstyle}

\usepackage{graphicx,amsmath}

\usepackage{astjnlabbrev-nature} 

\title{The diversity of quasars unified by accretion and orientation}

\author{Yue Shen$^{1,2,4}$, Luis C. Ho$^{2,3}$}

\newcommand{\etal}{et al.}
\newcommand{\hbeta}{H{$\beta$}}

\newcommand{\CIV}{C{\sevenrm IV}}

\newcommand{\HeII}{He{\sevenrm II}}

\def\MgII{Mg\,{\sc ii}}

\newcommand{\OII}{[O{\sevenrm\,II}]}

\newcommand{\NeV}{[Ne{\sevenrm\,V}]}
\def \OIII {[O\,{\sc iii}]}
\def \FeII {Fe\,{\sc ii}}

\def \NeV {[Ne\,{\sc v}]}
\def \OII {[O\,{\sc ii}]}
\def \NeIII {[Ne\,{\sc iii}]}
\newcommand{\OIIIa}{[O{\sevenrm\,III}]\,$\lambda$4959}
\newcommand{\OIIIb}{[O{\sevenrm\,III}]\,$\lambda$5007}
\newcommand{\OIIIab}{[O{\sevenrm\,III}]\,$\lambda\lambda$4959, 5007}

\newcommand{\SII}{[S{\sevenrm\,II}]}

   \font\sevenrm=cmr7 scaled 1000

\def\Rfe{R_{\textrm{\FeII}}}
\def\Mbh{$M_{\rm BH}$}

\begin{document}

\maketitle

\begin{affiliations}
 \item Carnegie Observatories, 813 Santa Barbara Street, Pasadena,
CA 91101, USA
 \item Kavli Institute for Astronomy and Astrophysics, Peking University, Beijing 100871, China 
 \item Department of Astronomy, School of Physics, Peking University, Beijing 100871, China
 \item Hubble Fellow
\end{affiliations}

\begin{abstract}
Quasars are rapidly accreting supermassive black holes at the center of massive galaxies. They display a broad range of properties across all wavelengths, reflecting the diversity in the physical conditions of the regions close to the central engine. These properties, however, are not random, but form well-defined trends. The dominant trend is known as Eigenvector 1, where many properties correlate with the strength of optical iron and \OIII\ emission\cite{Boroson_Green_1992,Wang_etal_1996,Sulentic_etal_2000a}. The main physical driver of Eigenvector 1 has long been suspected to be the quasar luminosity normalized by the mass of the hole (the Eddington ratio)\cite{Boroson_2002}, an important quantity of the black hole accretion process. But a definitive proof has been missing. Here we report an analysis of archival data that reveals that Eddington ratio indeed drives Eigenvector 1. We also find that orientation plays a significant role in determining the observed kinematics of the gas, implying a flattened, disklike geometry for the fast-moving clouds close to the hole.  Our results show that most of the diversity of quasar phenomenology can be unified with two simple quantities, Eddington ratio and orientation.

\end{abstract}

The optical and ultraviolet spectra of quasars show emission lines with a wide variety of strengths (equivalent width, EW) and velocity widths.  However, despite their great diversity in outward appearance, quasars, in fact, possess surprising regularity in their physical properties. A seminal principal component analysis\cite{Boroson_Green_1992} of 87 low-redshift broad-line quasars discovered that the main variance (Eigenvector 1, or EV1) in their optical properties arises from an anti-correlation between the strength of the narrow \OIII\ $\lambda$5007 and broad \FeII\ emission. Along with other properties that also correlate with \FeII\ strength\cite{Wang_etal_1996,Laor_etal_1997,Sulentic_etal_2000a}, these observations establish Eigenvector 1 as a physical sequence of broad-line quasar properties. In the two-dimensional plane of \FeII\ strength (measured by the ratio of \FeII\ EW within 4434-4684 \AA\ to broad \hbeta\ EW, $R_{\textrm{\FeII}}\equiv \textrm{EW}_\textrm{\FeII}/\textrm{EW}_\textrm{\hbeta}$) and the full-width-at-half-maximum of broad \hbeta\ (FWHM$_{\textrm{\hbeta}}$), Eigenvector 1 is defined as the horizontal trend with $\Rfe$, where the average \OIII\ strength and FWHM$_{\textrm{\hbeta}}$ decrease\cite{Boroson_Green_1992,Sulentic_etal_2000a}. Figure 1 shows the EV1 sequence for $\sim 20,000$ broad-line quasars drawn from the SDSS\cite{Schneider_etal_2010,Shen_etal_2011} (see Supplementary Information, SI, for details regarding the sample).

The statistics of the SDSS quasar sample allows us to divide the sample into bins of $R_{\textrm{\FeII}}$ and $\textrm{FWHM}_{\textrm{\hbeta}}$ (the gray grid in Figure 1) and study the average \OIII\ properties in each bin. Figure 2 shows the average \OIII\ line profiles in each bin, as a function of $L$, the quasar continuum luminosity measured at 5100\,\AA. In addition to the Eigenvector 1 sequence, the \OIII\ strength also decreases with $L$, following the Baldwin effect\cite{Stern_Laor_2013,Zhang_etal_2013} initially discovered for the broad \CIV\ line\cite{Baldwin_1977}. The \OIII\ profile can be decomposed into a core component, centered consistently at the systemic redshift, and a blueshifted, wing component. The core component strongly follows the EV1 and Baldwin trends, while the wing component only shows mild decrement with $L$ and $R_{\textrm{\FeII}}$ (SI and Extended Data Figures 1-2). This may suggest that the core component is mostly powered by photoionization from the quasar, while the wing component is excited by other mechanisms, such as shocks associated with outflows\cite{Dopita_Sutherland_1995}.  

\begin{figure}
\includegraphics[width=0.5\textwidth]{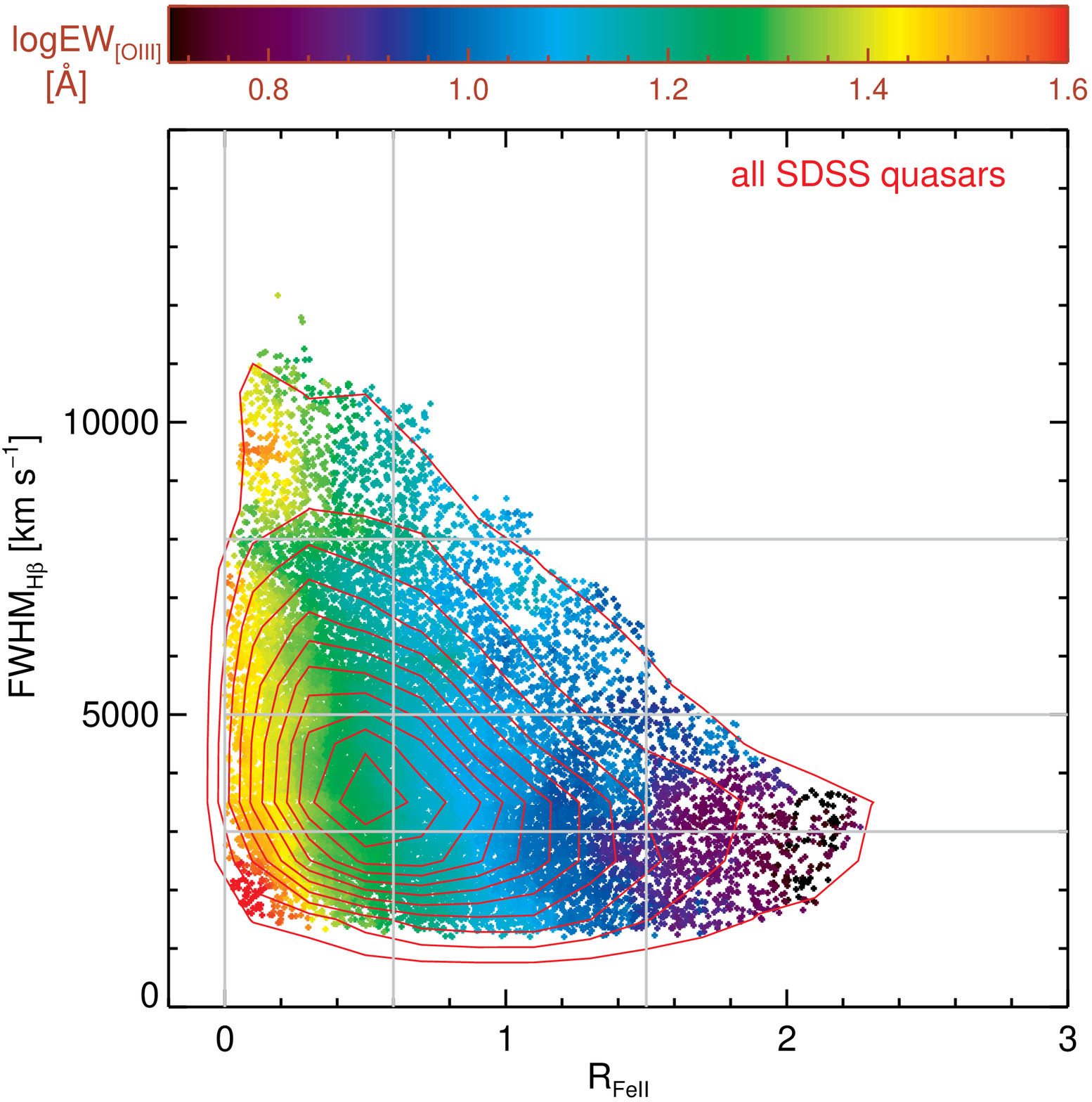}
\caption{\textbf{Distribution of quasars in the EV1 plane.} The horizontal axis is the relative \FeII\ strength, $\Rfe$, and the vertical axis is the broad \hbeta\ FWHM. The red contours show the distribution of our SDSS quasar sample, and the points show individual objects. We color-code the points by the \OIIIb\ strength, averaged over all nearby objects in a smoothing box of $\Delta\Rfe=0.2$ and $\Delta {\rm FWHM}_\textrm{\hbeta}=1000\,{\rm km\,s^{-1}}$. The EV1 sequence\cite{Boroson_Green_1992} is the systematic trend of decreasing \OIII\ strength with increasing $\Rfe$. The gray grid divides this plane into bins of FWHM$_\textrm{\hbeta}$ and $\Rfe$, in which we study the stacked spectral properties. } \label{fig:ev1_1}
\end{figure}

In addition to the strongest narrow \OIII\ lines, all other optical narrow forbidden lines (e.g., \NeV, \NeIII, \OII, \SII) show similar EV1 trends and Baldwin effect. Hot dust emission detected using WISE\cite{Wright_etal_2010}, presumably coming from a dusty torus\cite{Antonucci_93,Urry_Padovani_1995}, also increases with $R_{\textrm{\FeII}}$. In the SI (and Extended Data Figures 3-7) we summarize all updated and new observations that firmly establish the Eigenvector 1 sequence.

The \OIII\ emitting region is photoionized by the ionizing continuum from the accreting black hole (BH). But the Eigenvector 1 correlation of \OIII\ strength with $R_{\textrm{\FeII}}$ holds even when optical luminosity is fixed, as demonstrated in Figure 2. This suggests that another physical property of BH accretion changes with $\Rfe$, one that, in turn, affects the relative contribution in the ionizing part of the quasar continuum as seen by the narrow-line region. The most likely possibility is the BH mass $M_{\rm BH}$, or equivalently, the Eddington ratio $L/M_{\rm BH}$, since $L$ is fixed. The much less likely alternative would be that the \OIII\ narrow-line region changes as a function of $\Rfe$. Reverberation mapping (RM) studies of nearby active galactic nuclei (AGN)\cite{Peterson_etal_2004} have suggested that a virial estimate of $M_{\rm BH}$ may be derived by combining the broad-line region size $R_{\rm BLR}$ (measured from the time lag between continuum and emission-line variability) and the virial velocity of the line-emitting clouds estimated from the line width: $M_{\rm BH,vir}\propto R_{\rm BLR}{\rm FWHM}_\textrm{\hbeta}^2/G$. The average FWHM$_\textrm{\hbeta}$ does decrease by $\sim 0.2$ dex when $R_{\textrm{\FeII}}$ increases from 0 to 2, and this fact underlies the earlier suggestion that Eigenvector 1 is driven by Eddington ratio\cite{Laor_2000,Boroson_2002}.

A remarkable feature in Figures 1 and 2 is that the sequence is predominantly horizontal: there is little trend with FWHM$_{\textrm{\hbeta}}$ at fixed $R_{\textrm{\FeII}}$. The standard virial mass estimators\cite{Peterson_etal_2004,Vestergaard_Peterson_2006} would suggest that there is a strong vertical segregation in $M_{\rm BH}$, by a factor of a few in the vertical bins in Figure 1. If lower \Mbh\ (or higher Eddington ratio) leads to weaker \OIII\ as in the Eigenvector 1 relation (i.e., the horizontal trend), we should also see a vertical trend in Figure 1. The absence of such a trend suggests that there is substantial scatter between FWHM$_{\textrm{\hbeta}}$ and the actual virial velocity, and the vertical spread in FWHM$_{\textrm{\hbeta}}$ in the EV1 plane largely does not track the spread in true BH masses. 

We propose, instead, that the sequence in $R_{\textrm{\FeII}}$ is driven by \Mbh; but the dispersion in FWHM$_{\textrm{\hbeta}}$ at fixed $R_{\textrm{\FeII}}$ is due to an orientation effect, as expected in a flattened broad-line region geometry. 
We first demonstrate that the average \Mbh\ indeed decreases with $R_{\textrm{\FeII}}$ for our quasar sample. We achieve this by measuring the clustering of SDSS quasars with low and high $R_{\textrm{\FeII}}$ values. In the hierarchical clustering Universe, more massive galaxies (which contain more massive BHs) form in rarer density peaks and are more strongly clustered\cite{BBKS}. We therefore expect quasars with larger $R_{\textrm{\FeII}}$ are less strongly clustered. This exercise, however, has a stringent requirement on sample statistics, and is not possible until now. Here we take advantage of the largest spectroscopic sample of galaxies from SDSS-III\cite{Ahn_etal_2013}, and use the much larger (by a factor of $\sim 40$) galaxy sample to cross-correlate\cite{Shen_etal_2013} with our quasar sample at $z\sim 0.5$ to substantially improve the clustering measurements. The resulting cross-correlation functions are shown in the left panel of Figure 3, for the two quasar subsamples divided at the median $R_{\textrm{\FeII}}$. A significant clustering difference is detected at $3.48\sigma$: quasars with larger $R_{\textrm{\FeII}}$ indeed are less strongly clustered, confirming that they have on average lower \Mbh. 

\begin{figure}
\includegraphics[width=0.5\textwidth]{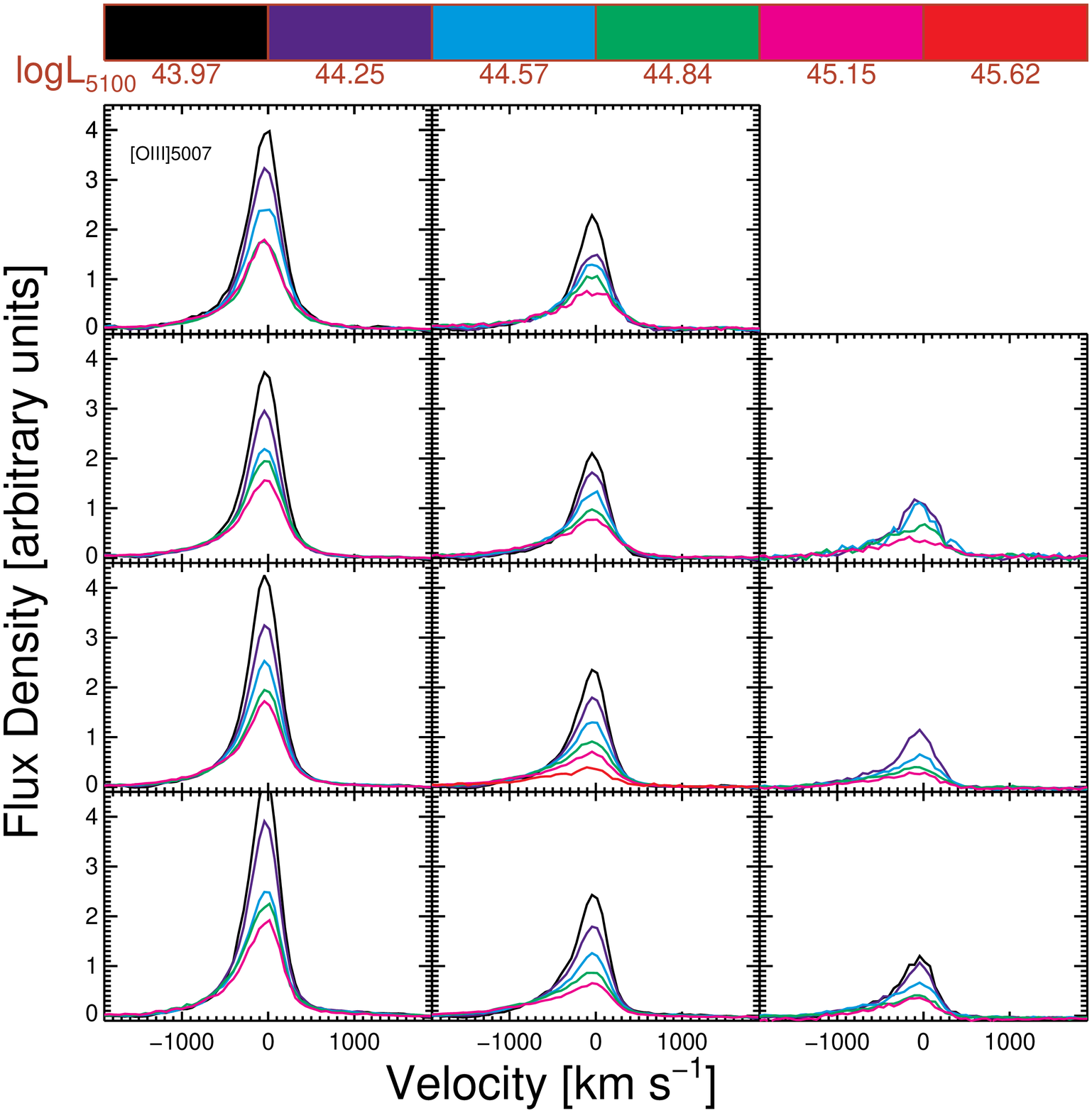}
\caption{\textbf{Average \OIII\ profiles in the EV1 plane.} Each panel shows the stacked \OIIIb\ line of quasars in the $\Rfe-$FWHM$_\textrm{\hbeta}$ bins defined by the gray grid in Figure 1 (in the same layout). $\Rfe$ increases from left to right, and FWHM$_\textrm{\hbeta}$ increases from bottom to top. In each bin we further divided the quasars into different luminosity bins using the measured $L_{5100}$ continuum luminosities. We have normalized the line fluxes by the (host-corrected) average quasar continuum luminosity $L_{5100}$ for each stacking subset; hence, these stacked lines reflect the relative \OIII\ strength among different samples. In addition to the decrease of \OIII\ strength when $\Rfe$ increases (i.e., Figure 1), we also observe a decrease in \OIII\ strength with increasing luminosity\cite{Stern_Laor_2013,Zhang_etal_2013}. The \OIII\ profile is in generally asymmetric, with a blueshifted wing, whose relative contribution to the total profile increases when $\Rfe$ or luminosity increases. }\label{fig:ev1_oiii_vel}
\end{figure}

\begin{figure}
 \centering
 \includegraphics[width=0.48\textwidth]{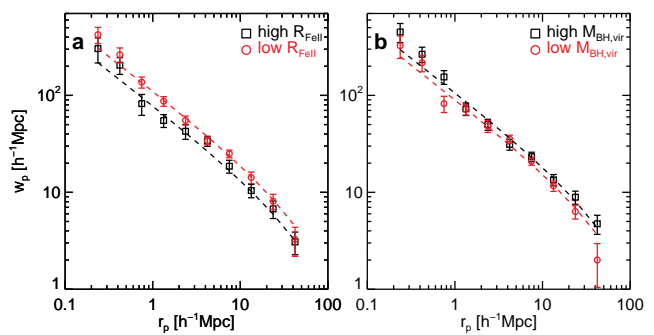}
 \caption{\textbf{Cross-correlation functions between different quasar subsamples and a galaxy sample.} {\em Left:} difference in the clustering strength when the quasar sample is divided by the median $\Rfe$. A significant difference ($3.48\sigma$) is detected: quasars with stronger $\Rfe$ are less strongly clustered, indicating they have on average smaller BH masses. {\em Right:} difference in the clustering strength when the quasar sample is divided by the virial BH mass estimates based on FWHM$_\textrm{\hbeta}$. No significant difference ($1.64\sigma$) is detected, indicating there is substantial overlap in the actual BH masses between the two subsamples due to the uncertainties in these FWHM-based virial BH masses. Orientation-induced FWHM$_\textrm{\hbeta}$ dispersion can naturally lead to such uncertainties. Error bars are measurement errors estimated with jackknife resampling (SI). 
 }
 \label{fig:clustering} \label{fig:clustering}
\end{figure}

In the Eigenvector 1 plane (Figure 1), the distribution in FWHM$_{\textrm{\hbeta}}$ at fixed $R_{\textrm{\FeII}}$ is roughly lognormal, with mean value decreasing with $R_{\textrm{\FeII}}$ and a dispersion of $\sim 0.2$ dex (Extended Data Figure 8). We argued above that this dispersion is largely orientation-induced FWHM variations in the case of a flattened broad-line region geometry. For a small subset of quasars that are radio-loud ($\sim 10\%$ of the population), it is possible to infer the orientation of the accretion disk, and by extension, the broad-line region, using resolved radio morphology to deduce the orientation of the jet. Such studies\cite{Wills_Browne_1986,Runnoe_etal_2012} show that high-inclination (more edge-on) broad-line radio quasars have on average larger FWHM$_{\textrm{\hbeta}}$, in accordance with the orientation hypothesis. Below we perform a different test for the more general radio-quiet quasar population, and we provide further evidence to support this argument in the SI and Extended Data Figures 9-10.

We compile a sample of 29 low-redshift AGNs with literature broad-line region size measurements from RM\cite{Peterson_etal_2004}, host stellar velocity dispersion ($\sigma_*$) measurements\cite{Park_etal_2012}, and optical spectroscopy\cite{Marziani_etal_2003}. We use the well-established local $M_{\rm BH}-\sigma_*$ relation\cite{KH13} to independently estimate BH masses for the 29 AGNs. We supplement the 29 local AGNs with a sample of $\sim 600$ SDSS AGNs\cite{Shen_etal_2008x}, where the host stellar velocity dispersion was estimated from spectral decomposition of the SDSS spectrum into AGN and host galaxy components, and the broad-line region size $R_{\rm BLR}$ was estimated using the tight correlation between $R_{\rm BLR}$ and the AGN luminosity found in RM studies\cite{Bentz_etal_2009}. We can then define a virial coefficient, $f\equiv GM_{\rm BH}/(R_{\rm BLR} {\rm FWHM}^2_{\textrm{\hbeta}})$. At a given $M_{\rm BH}$, $f$ should not depend on FWHM$_{\textrm{\hbeta}}$, if the latter is a faithful indicator of the broad-line region virial velocity. However, if FWHM$_{\textrm{\hbeta}}$ is orientation-dependent as suggested above, $f$ will be anti-correlated with FWHM$_{\textrm{\hbeta}}$. 

\begin{figure}
 \centering
 \includegraphics[width=0.48\textwidth]{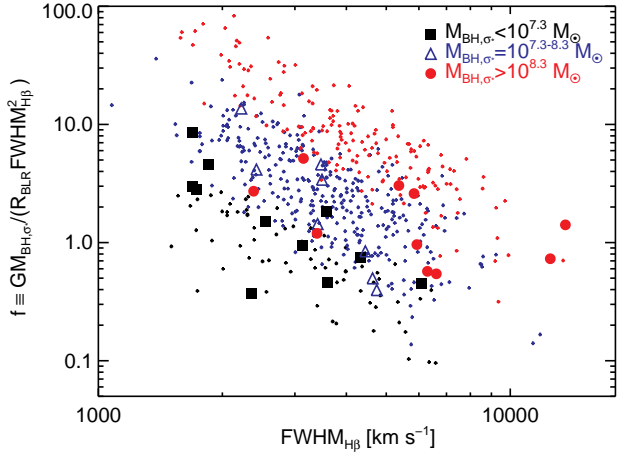}
 \caption{\textbf{The effect of orientation on FWHM$_\textrm{\hbeta}$.} The large symbols represent the 29 low-redshift AGNs that have both reverberation mapping data and host stellar velocity dispersion ($\sigma_*$) measurements. The small symbols represent a low-redshift SDSS AGN sample\cite{Shen_etal_2008x} with $\sigma_*$ and AGN spectral measurements based on spectral decomposition. We use the stellar velocity dispersion measurements and the local relation between BH mass and $\sigma_*$ from inactive galaxies\cite{KH13} to estimate the BH mass ($M_{\rm BH,\sigma_*}$) in these objects. We also estimate the average broad-line region size ($R_{\rm BLR}=c\tau$, where $c$ is the speed of light, and $\tau$ is the measured RM lag) in these objects, either from direct RM measurements, or by using the tight correlation between the broad-line region size and AGN luminosity\cite{Bentz_etal_2009}. The ratio of $M_{\rm BH,\sigma_*}$ to the product of $R_{\rm BLR}{\rm FWHM}_\textrm{\hbeta}^2/G$ (i.e., the virial coefficient $f$) is plotted as a function of FWHM$_\textrm{\hbeta}$, for different $M_{\rm BH,\sigma_*}$ values. The strong trends of $f$ with FWHM$_\textrm{\hbeta}$ at a given $M_{\rm BH,\sigma_*}$ suggest that the dispersion in FWHM$_\textrm{\hbeta}$ does not reflect the underlying virial velocity of the broad-line region gas, and tend to bias the BH mass estimates. This is in line with the fact that there is little vertical trend in the \OIII\ strength in the EV1 plane (Figure 1).  
 }
 \label{fig:rm} \label{fig:rm}
\end{figure}

Indeed, there is a strong dependence of $f$ on FWHM$_{\textrm{\hbeta}}$ at fixed $M_{\rm BH}$, shown in Figure 4, consistent with the orientation hypothesis. A direct consequence is that the standard virial BH mass estimates using FWHM$_{\textrm{\hbeta}}$ are subject to a significant uncertainty ($\sim 0.4$ dex) due to this orientation dependence. To test this, we perform the same cross-correlation analysis as above, but for quasar subsamples divided by their virial BH mass estimates based on FWHM$_{\textrm{\hbeta}}$. The results are shown in Figure 3: there is no significant detection (1.64$\sigma$) in the clustering difference between the two quasar subsamples. This is in accordance with there being substantial overlap in the true BH masses between the two subsamples, due to the uncertainty in virial BH mass estimates induced by using FWHM$_{\textrm{\hbeta}}$. The division by $R_{\textrm{\FeII}}$ provides a cleaner separation of high-mass BHs from low-mass ones in our sample. 



The collective evidence from this work leads to a rather simple interpretation of the observed main sequence of quasars (Figure 1): the average Eddington ratio increases from left to right, and the dispersion in FWHM$_\textrm{\hbeta}$ at fixed $R_{\textrm{\FeII}}$ is largely an orientation effect. The many physical quasar properties correlated with Eigenvector 1 are then unified as driven by changes in the average Eddington ratio of the BH accretion. While we do not discuss any physical model here, we suggest that the trends with the Eddington ratio are most likely caused by the systematic change in the shape of the accretion disk continuum and its interplay with the ambient emitting regions, which may in turn change the ionizing continuum as seen by the emission-line regions by modifying the structure of the accretion flow.


\begin{addendum}
 \item Support for the work of Y.S. was provided by NASA through Hubble Fellowship grant number HST-HF-51314.01, awarded by the Space Telescope Science Institute, which is operated by the Association of Universities for Research in Astronomy, Inc., for NASA, under contract NAS 5-26555. L.C.H. acknowledges support from the Kavli Foundation, Peking University, and the Chinese Academy of Science through grant No. XDB09030102 (Emergence of Cosmological Structures) from the Strategic Priority Research Program. This work makes extensive use of SDSS-I/II and SDSS-III data. The SDSS-I/II Web Site is http://www.sdss.org/. The SDSS-III Web Site is http://www.sdss3.org/.
 \item[Author Contributions] Y.S. and L.C.H. co-developed the idea; Y.S. performed the measurements and analysis; both authors contributed to the interpretation and manuscript writing. 
 \item[Competing Interests] The authors declare that they have no
competing financial interests.
 \item[Correspondence] Correspondence and requests for materials
should be addressed to Y.S. \\
(email: yshen@obs.carnegiescience.edu).
\end{addendum}



\clearpage


\begin{figure*}[!h]
\renewcommand\thefigure{E1}
 \centering
 \includegraphics[width=0.48\textwidth]{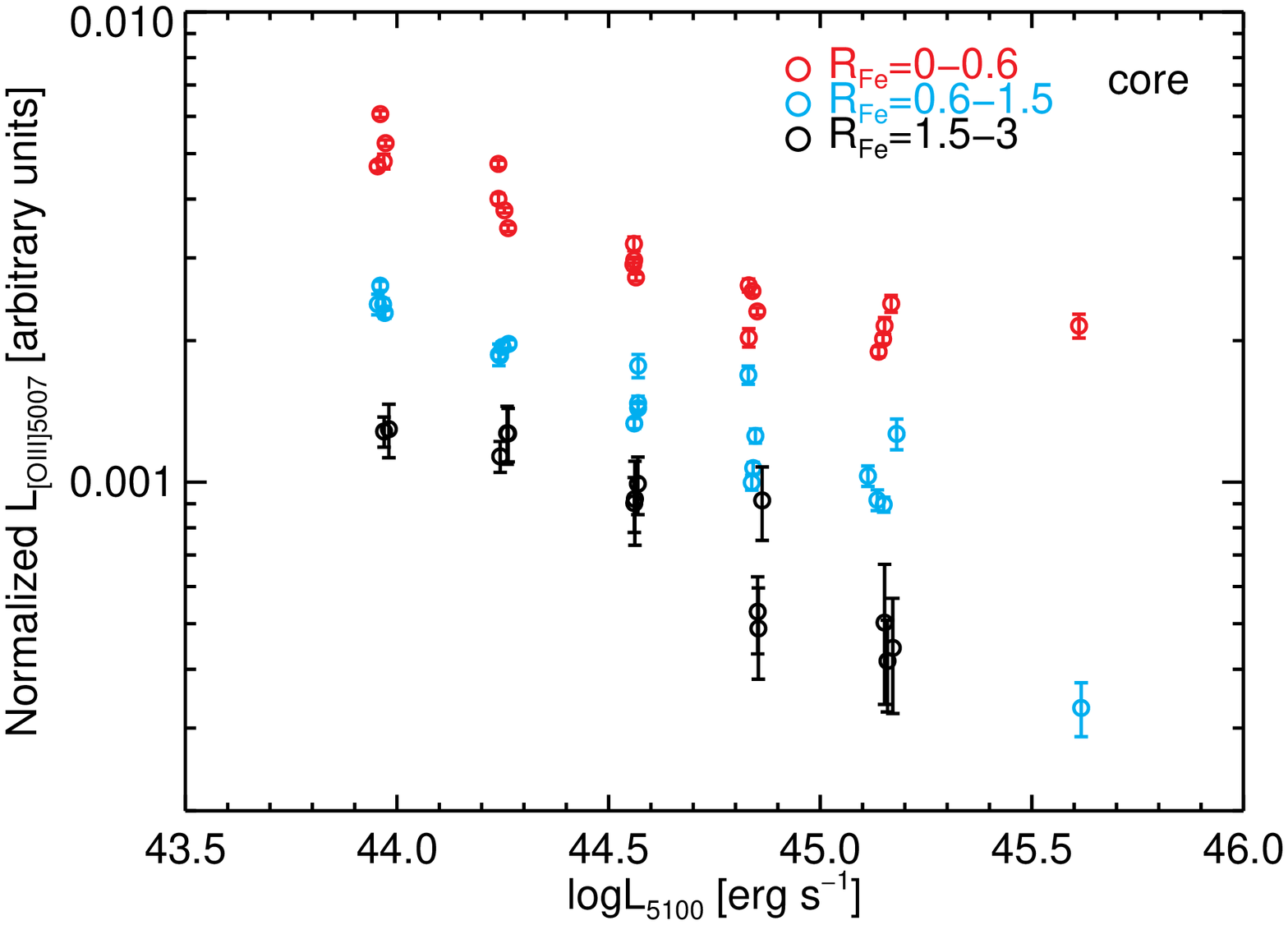}
 \includegraphics[width=0.48\textwidth]{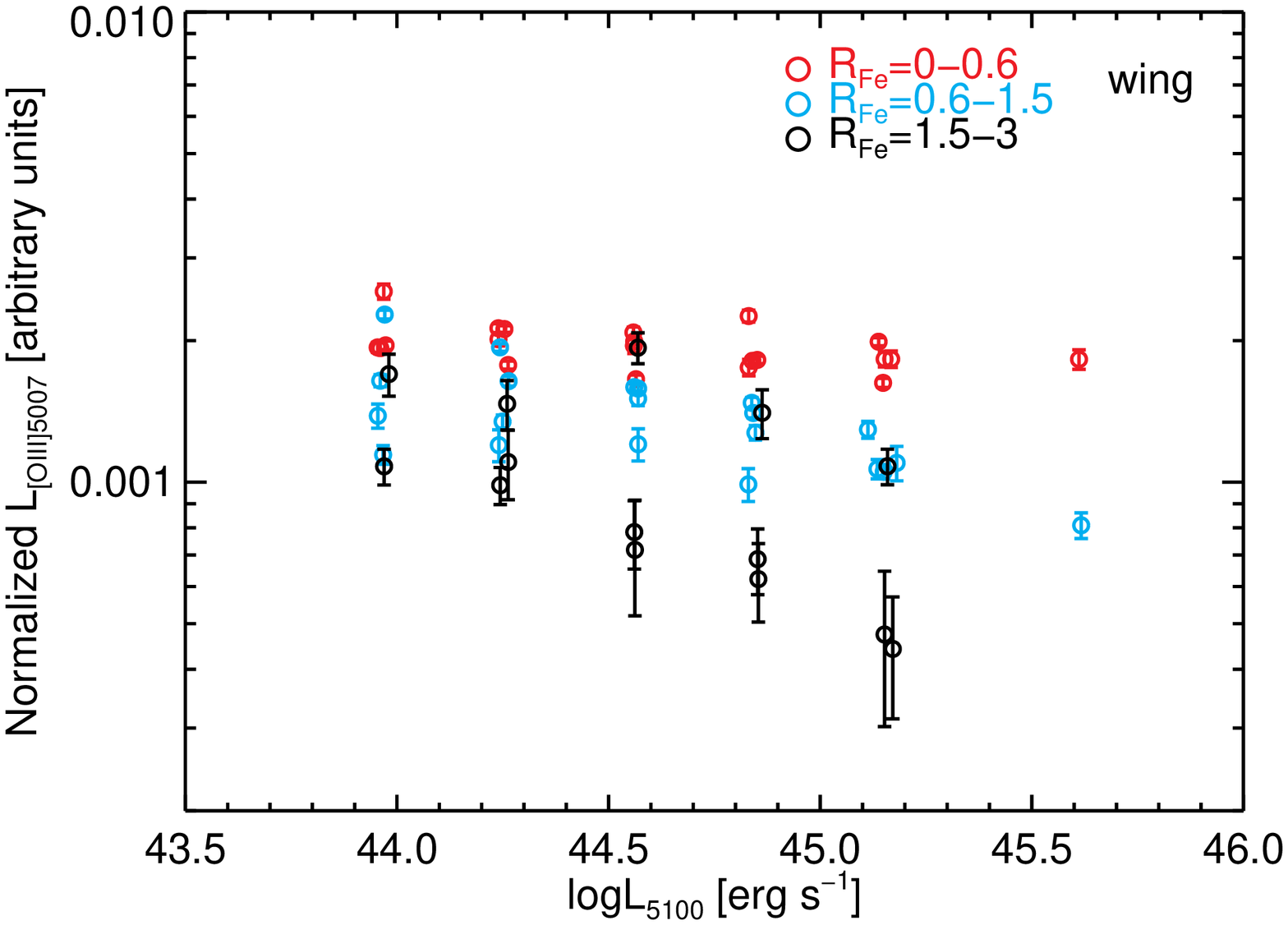}
 \caption{\textbf{Decomposed \OIIIb\ luminosity.} The left panel shows the core component and the right panel shows the wing component, for each composite spectrum shown in Figure\ 2. Measurement errors are estimated using Monte Carlo trials of mock spectra generated using the estimated flux error arrays of the coadded spectra. Both luminosities are normalized to the quasar continuum luminosity $L_{5100}$, hence reflect the strength of \OIII. The core \OIII\ shows a prominent anti-correlation with both $L_{5100}$ and $\Rfe$, while the wing \OIII\ shows weaker anti-correlations with $L_{5100}$ and $\Rfe$. For both \OIII\ components there is no correlation with FWHM$_{\textrm{\hbeta}}$, as evidenced in Figures\ 1 and 2. The Baldwin effect and EV1 correlation for \OIII\ shown in Figure\ 1 and Figure\ 2 are then primarily associated with the core \OIII\ component. The difference between the core and wing \OIII\ component may suggest different excitation mechanisms for both components. 
 }
 \label{fig:ev1_oiii_lum_decomp}
\end{figure*}

\begin{figure*} 
\renewcommand\thefigure{E2}
 \centering
 \includegraphics[width=0.48\textwidth]{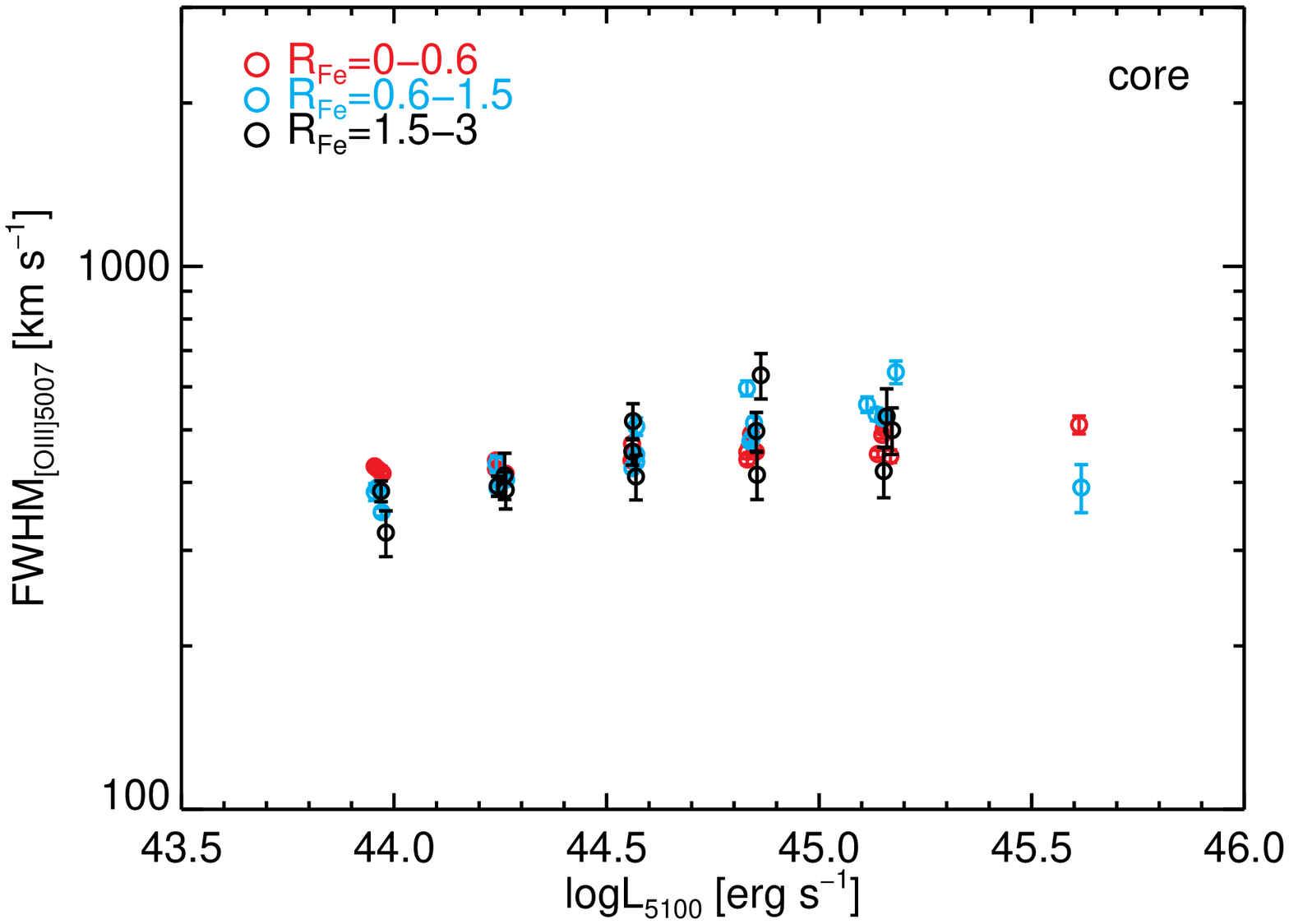}
 \includegraphics[width=0.48\textwidth]{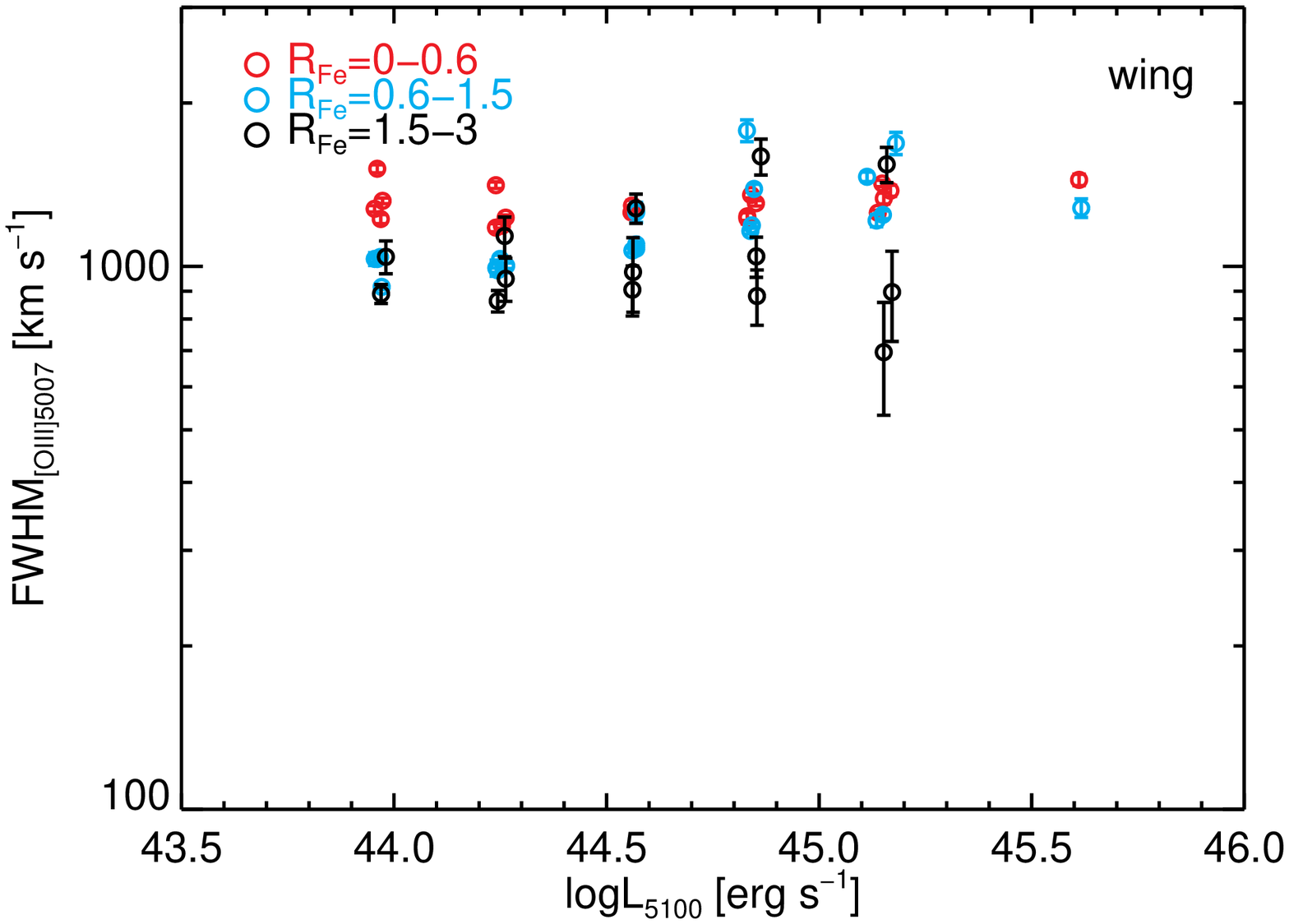}
 \includegraphics[width=0.48\textwidth]{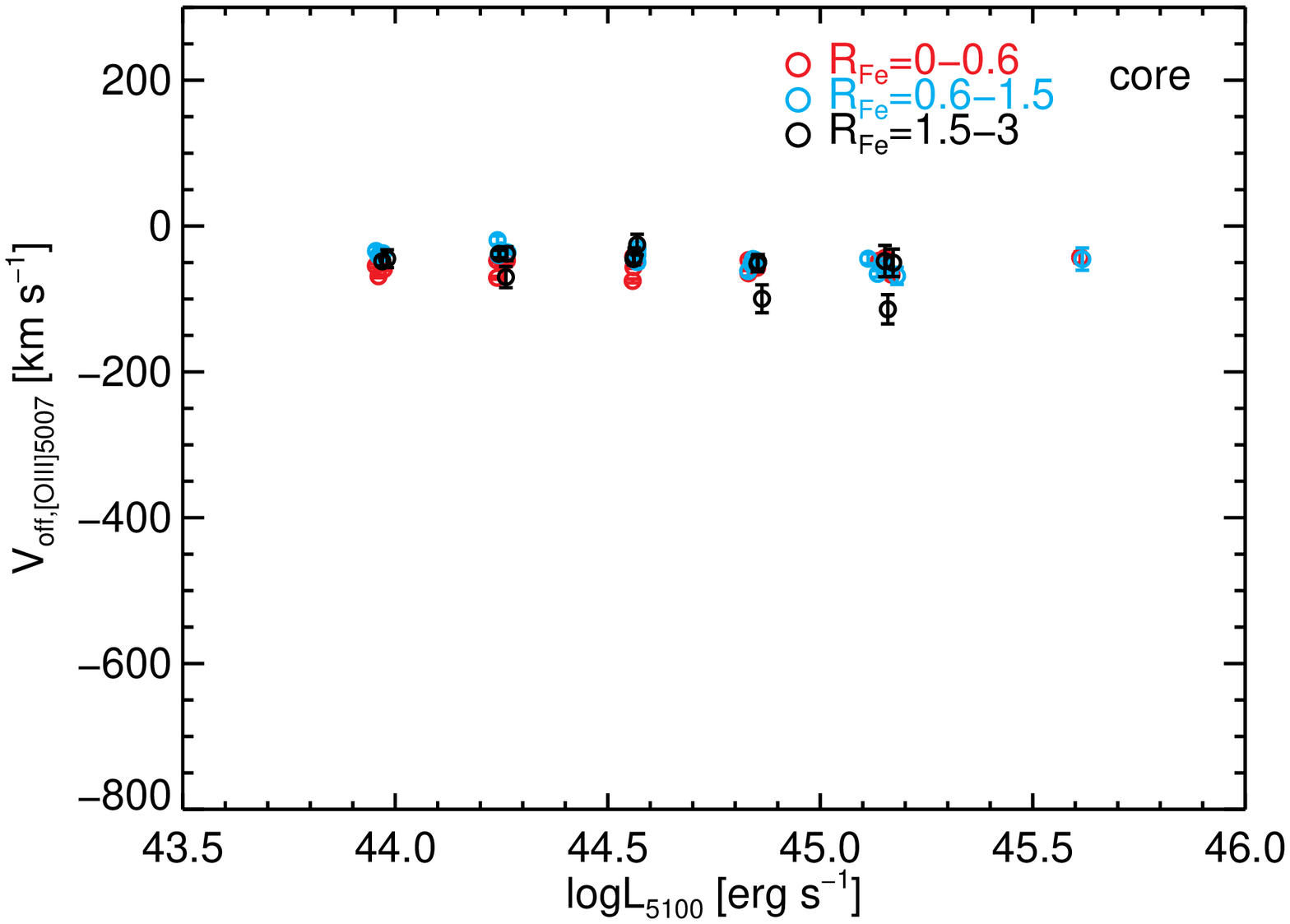}
 \includegraphics[width=0.48\textwidth]{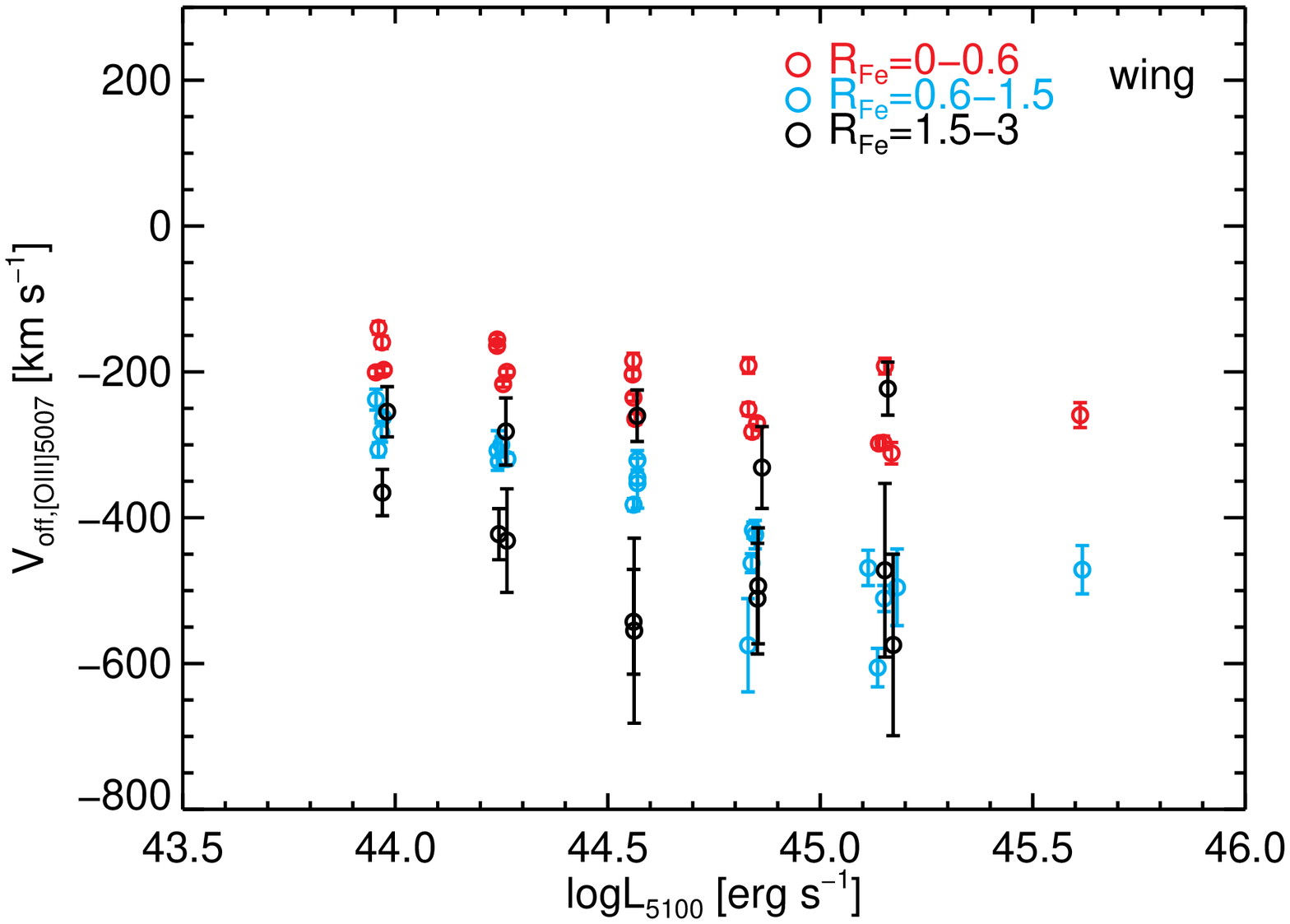}
 \caption{\textbf{Kinematic properties of the decomposed core and wing \OIII\ components.} Measurement errors are estimated using Monte Carlo trials of mock spectra generated using the estimated flux error arrays of the coadded spectra. The most significant correlations are the correlation between luminosity and the core \OIII\ FWHM, and the correlations between the wing \OIII\ blueshift and luminosity/$\Rfe$. The former correlation is consistent with the scenario that more luminous quasars are on average hosted by more massive galaxies with deeper potential well, hence larger core \OIII\ width. The latter correlations are consistent with the scenario that the wing \OIII\ component is associated with outflows. 
 }
 \label{fig:ev1_oiii_kin_decomp}
\end{figure*}

\begin{figure*}
\renewcommand\thefigure{E3}
 \centering
 \includegraphics[width=0.48\textwidth]{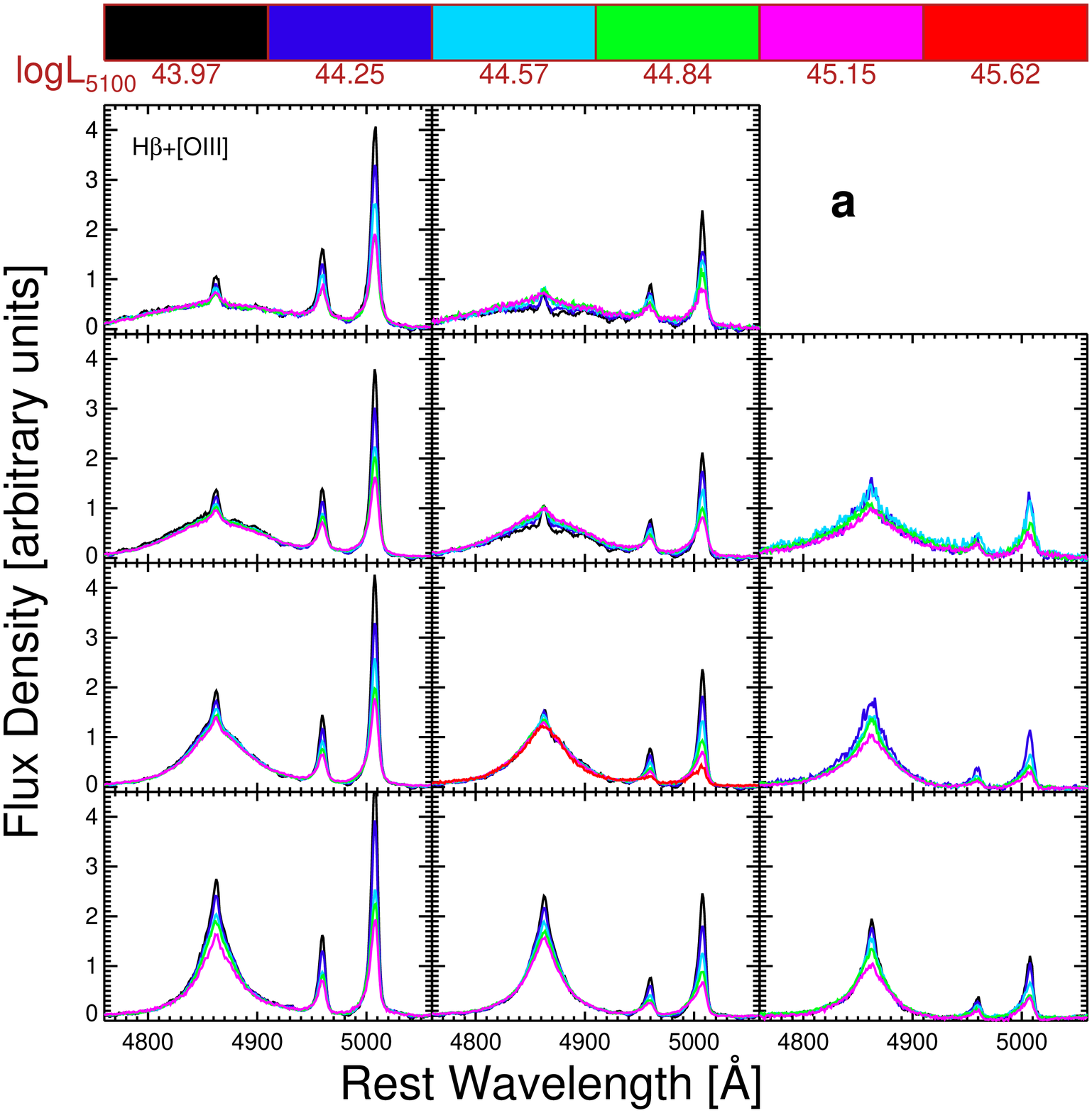}
 \includegraphics[width=0.48\textwidth]{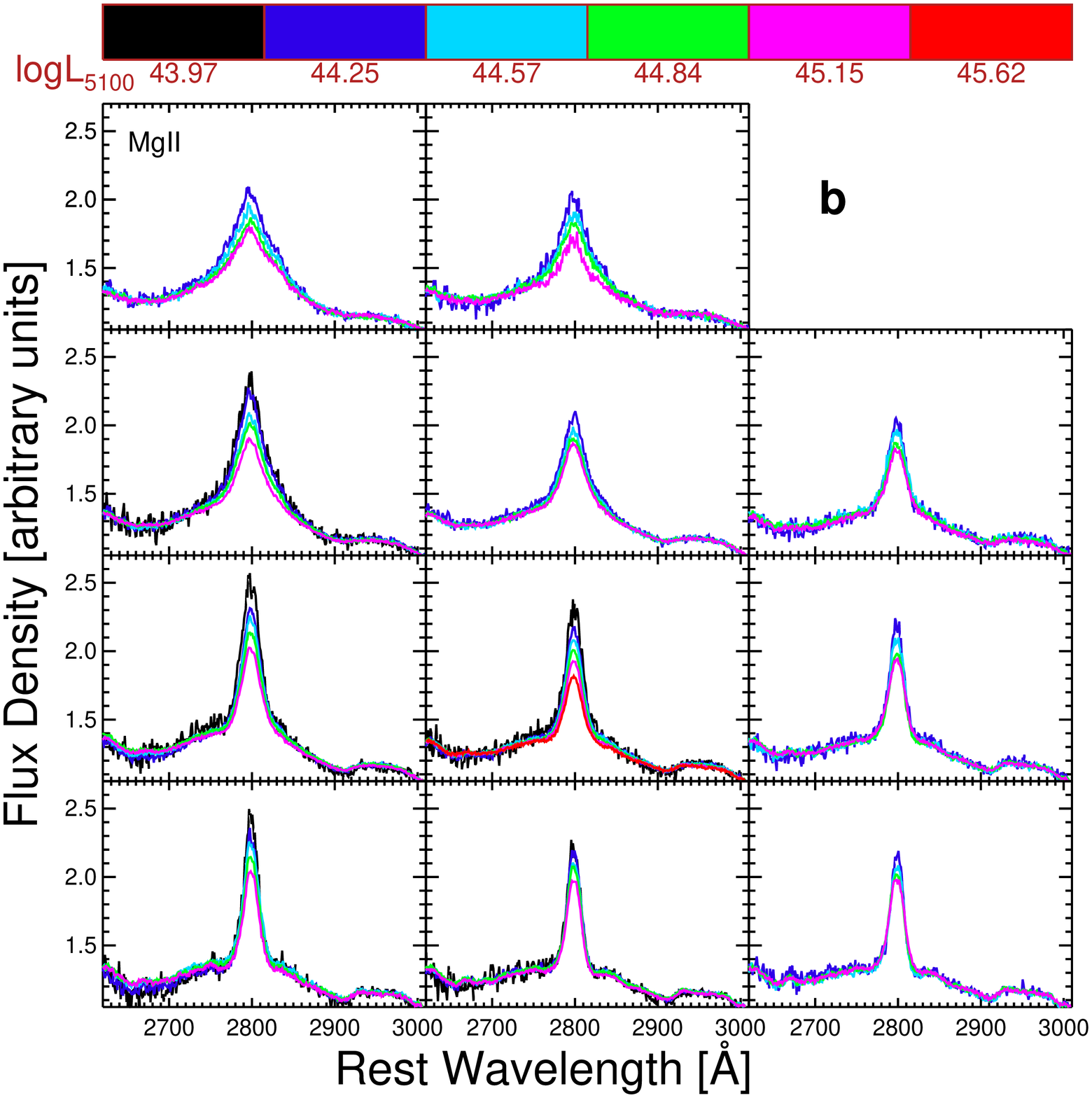}
 \includegraphics[width=0.48\textwidth]{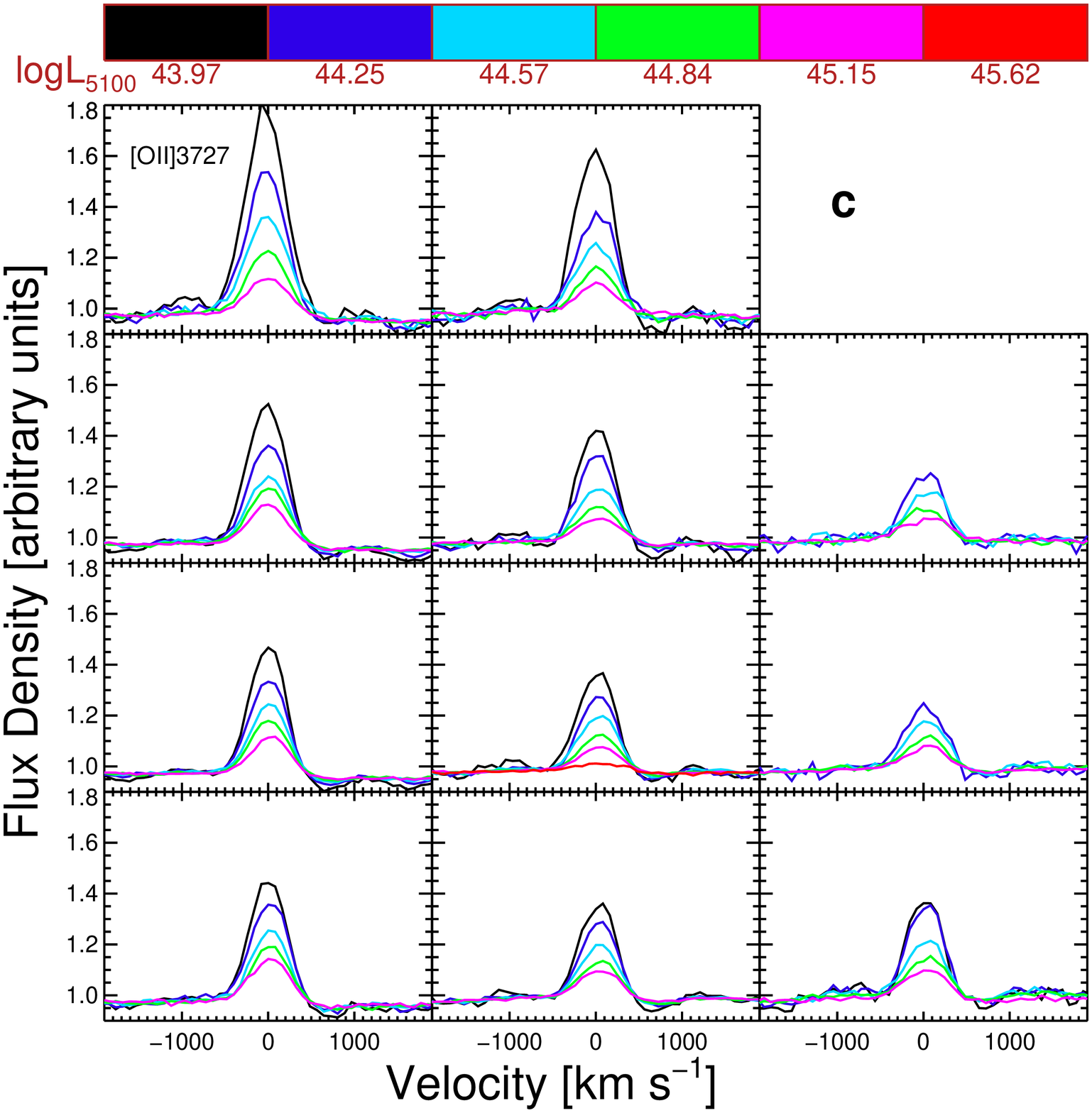}
 \includegraphics[width=0.48\textwidth]{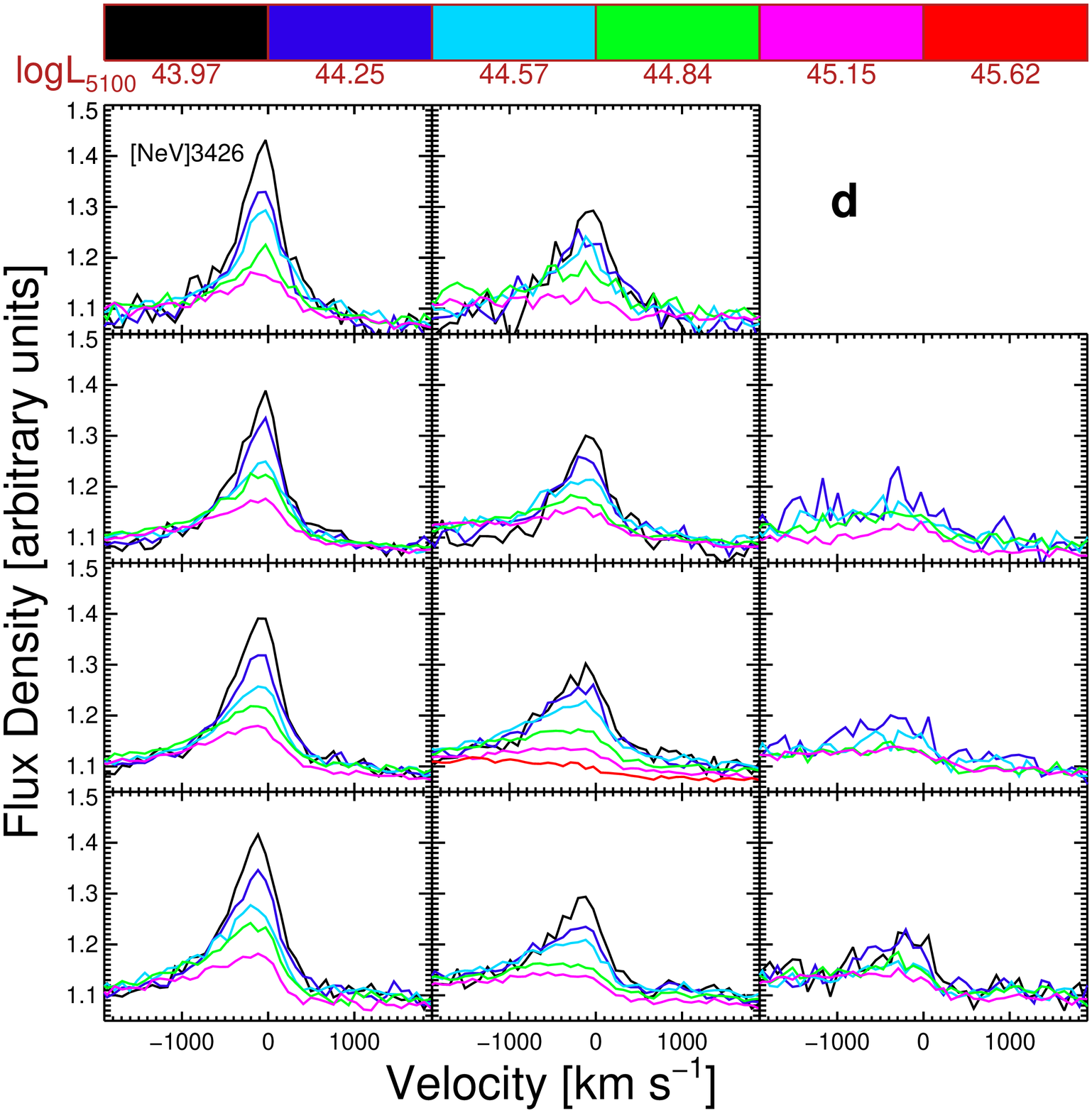}
 \caption{\textbf{Composite SDSS quasar spectra for several other lines in the same $\Rfe$-FWHM$_\textrm{\hbeta}$ bins defined in Figure\ 1.} As in Figure\ 2, each composite spectrum has been normalized by the continuum such that the integrated line intensity reflects the strength of the line. The composite spectra for the \hbeta\ region are generated using the pseudo-continuum-subtracted spectra, while for the other three lines (\MgII, \OII, and \NeV) the composite spectra are the median spectrum created using the full SDSS spectra and normalized at a nearby continuum window. 
 }
 \label{fig:ev1_other_line}
\end{figure*}

\begin{figure*}
\renewcommand\thefigure{E4}
 \centering
 \includegraphics[width=0.48\textwidth]{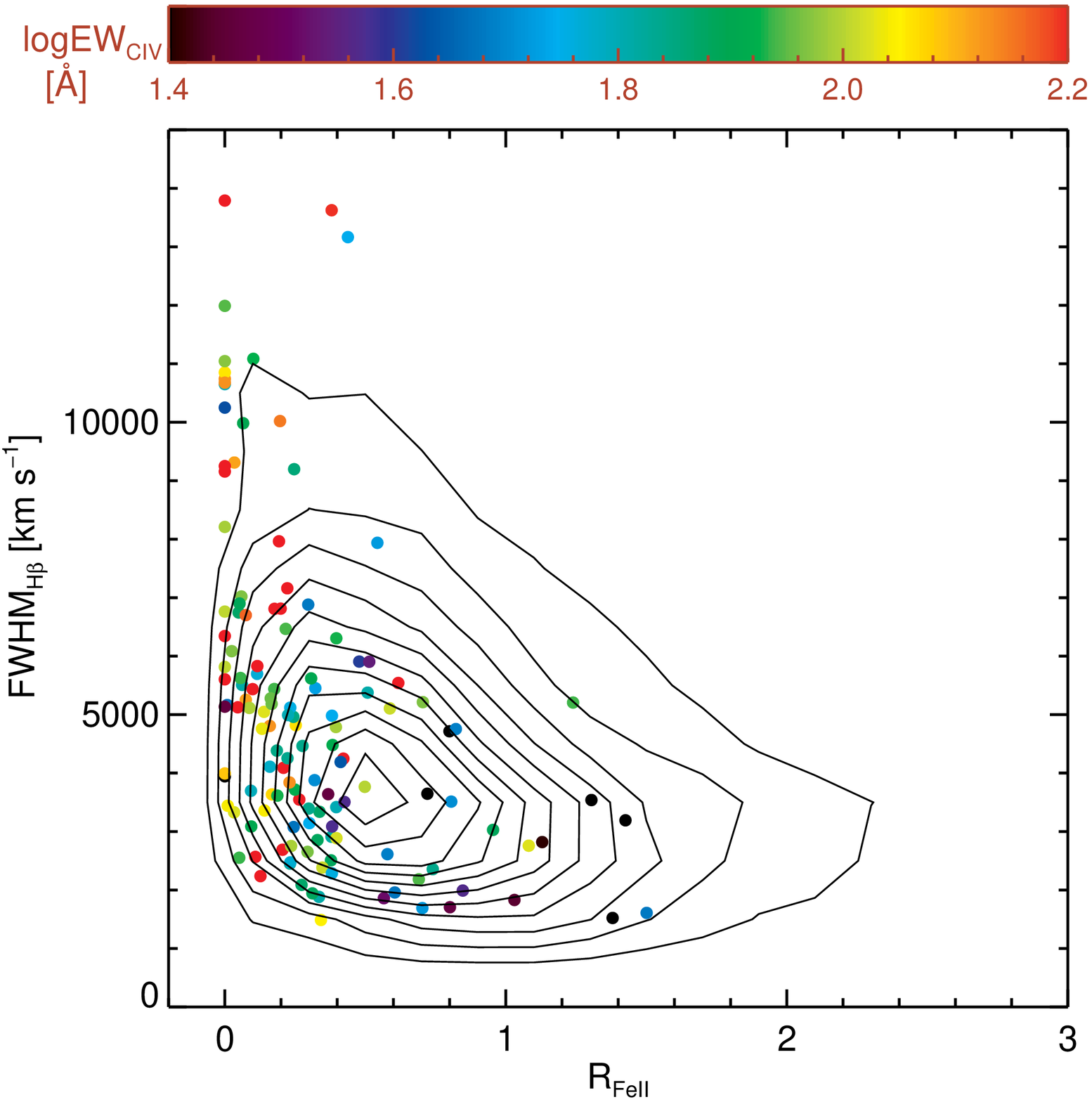}
 \caption{\textbf{Distribution in the EV1 plane in terms of \CIV\ properties.} A sample of low-redshift quasars with both \hbeta\ and \CIV\ measurements is shown, color-coded by the \CIV\ strength. A clear trend of decreasing \CIV\ strength with $\Rfe$ is seen, consistent with that seen for the other forbidden lines. The typical measurement uncertainties in \CIV\ EW is $\sim 7\%$ (relative), hence is negligible compared to the strong EV1 trend observed.
 }
 \label{fig:civ}
\end{figure*}

\begin{figure*}
\renewcommand\thefigure{E5}
 \centering
 \includegraphics[width=0.48\textwidth]{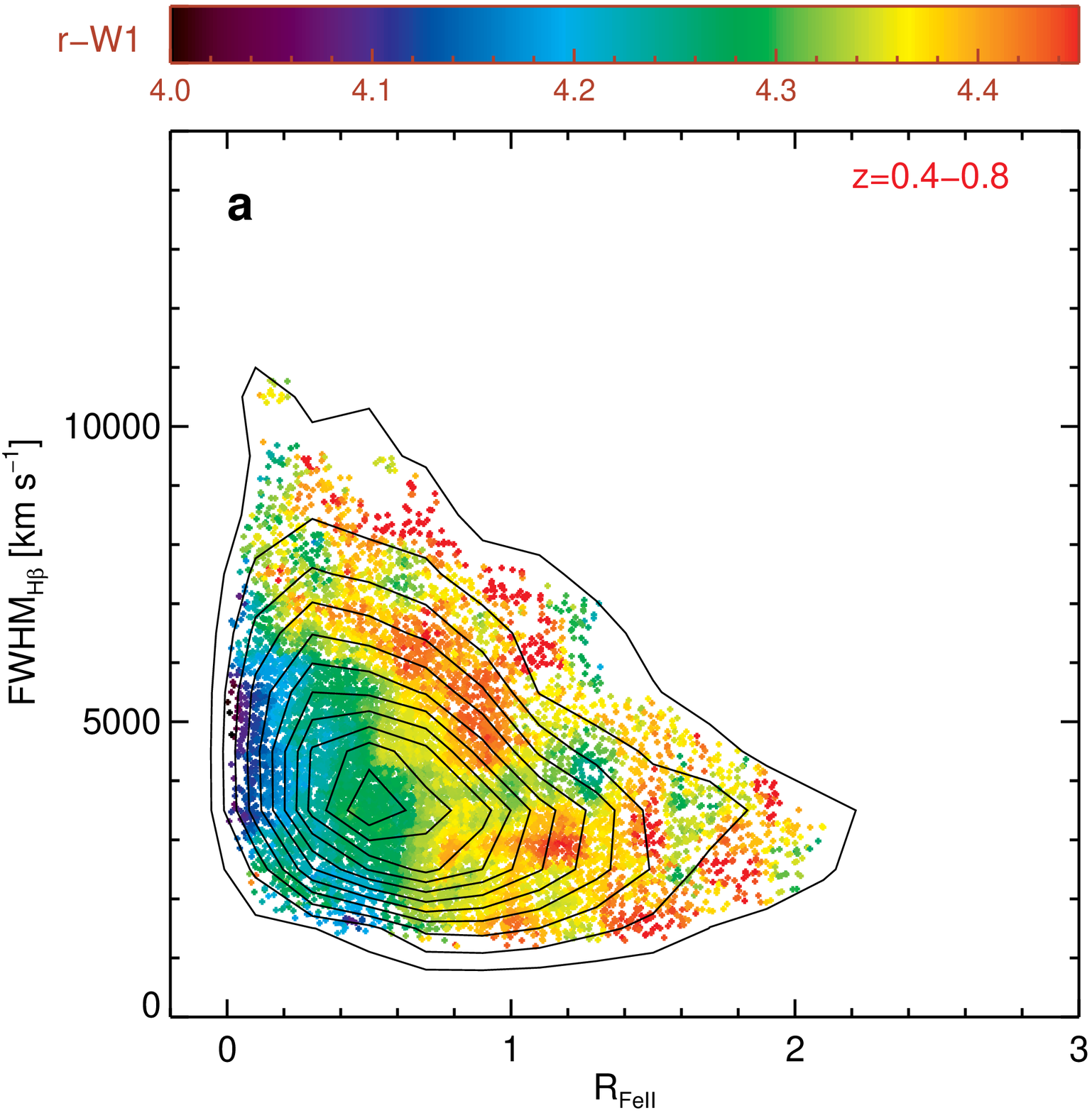}
 \includegraphics[width=0.48\textwidth]{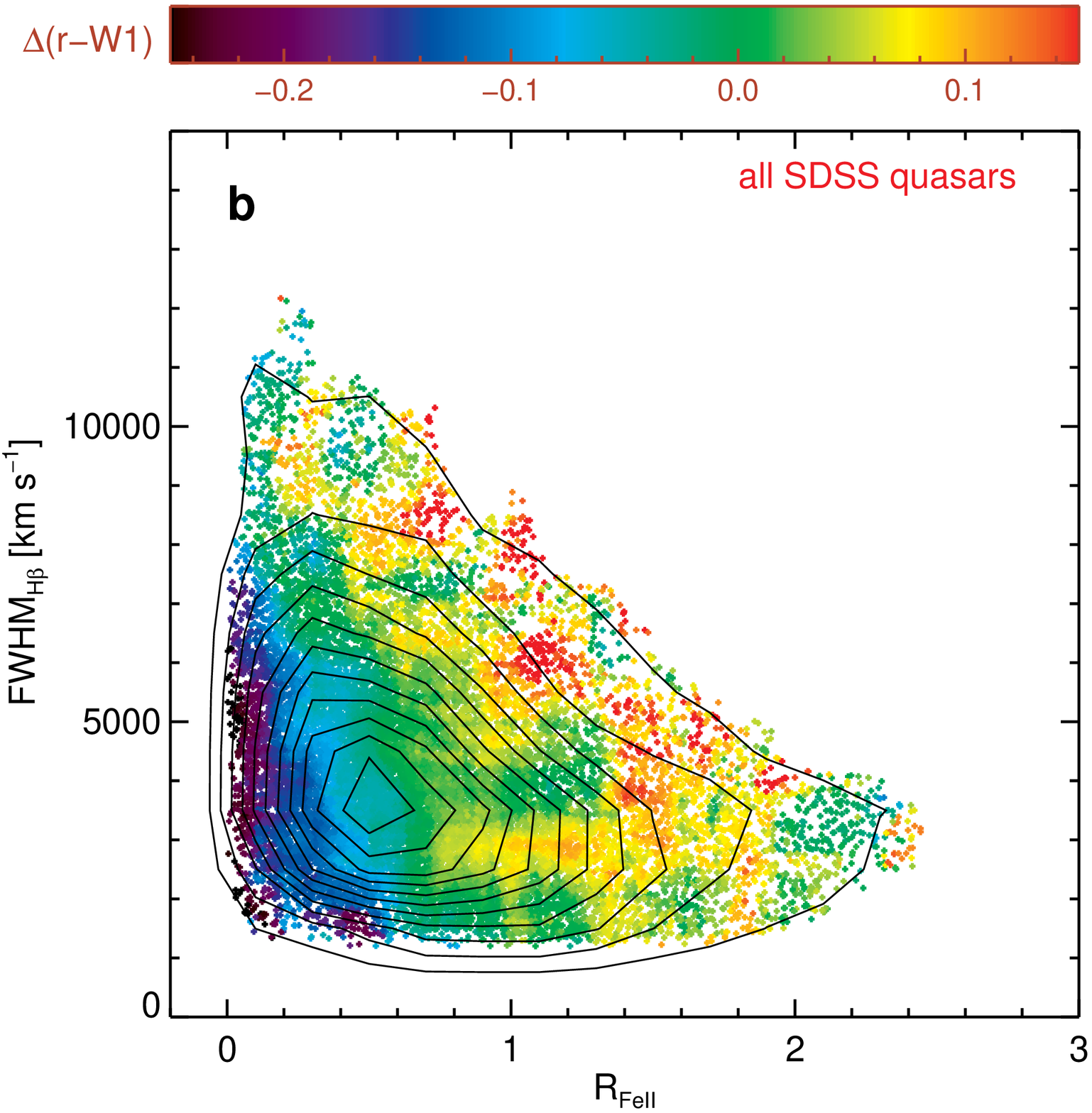}
 \caption{\textbf{Distributions of SDSS quasars in the EV1 plane in terms of the optical-infrared ($r-W1$) color.} The left panel shows $r-W1$ for quasars with $0.4<z<0.8$, for which the band shifting effect is small. We see a trend of increasing mid-infrared emission relative to optical emission with increasing $\Rfe$.  The right panel shows a similar result, using the excess color, $\Delta (r-W1)$, the deviation of $r-W1$ color from the mean color at each redshift. The usage of $\Delta (r-W1)$ removes the redshift dependence of colors, and we can apply this to all quasars in our sample. This test suggests that the torus emission is enhanced in quasars with larger $\Rfe$. Since we have argued that $\Rfe$ is a good indicator for the Eddington ratio, quasars with higher Eddington ratios have stronger torus emission, which may have implications for the formation mechanism of the dusty torus. 
 }
 \label{fig:wise}
\end{figure*}

\begin{figure*}
\renewcommand\thefigure{E6}
 \centering
 \includegraphics[width=0.48\textwidth]{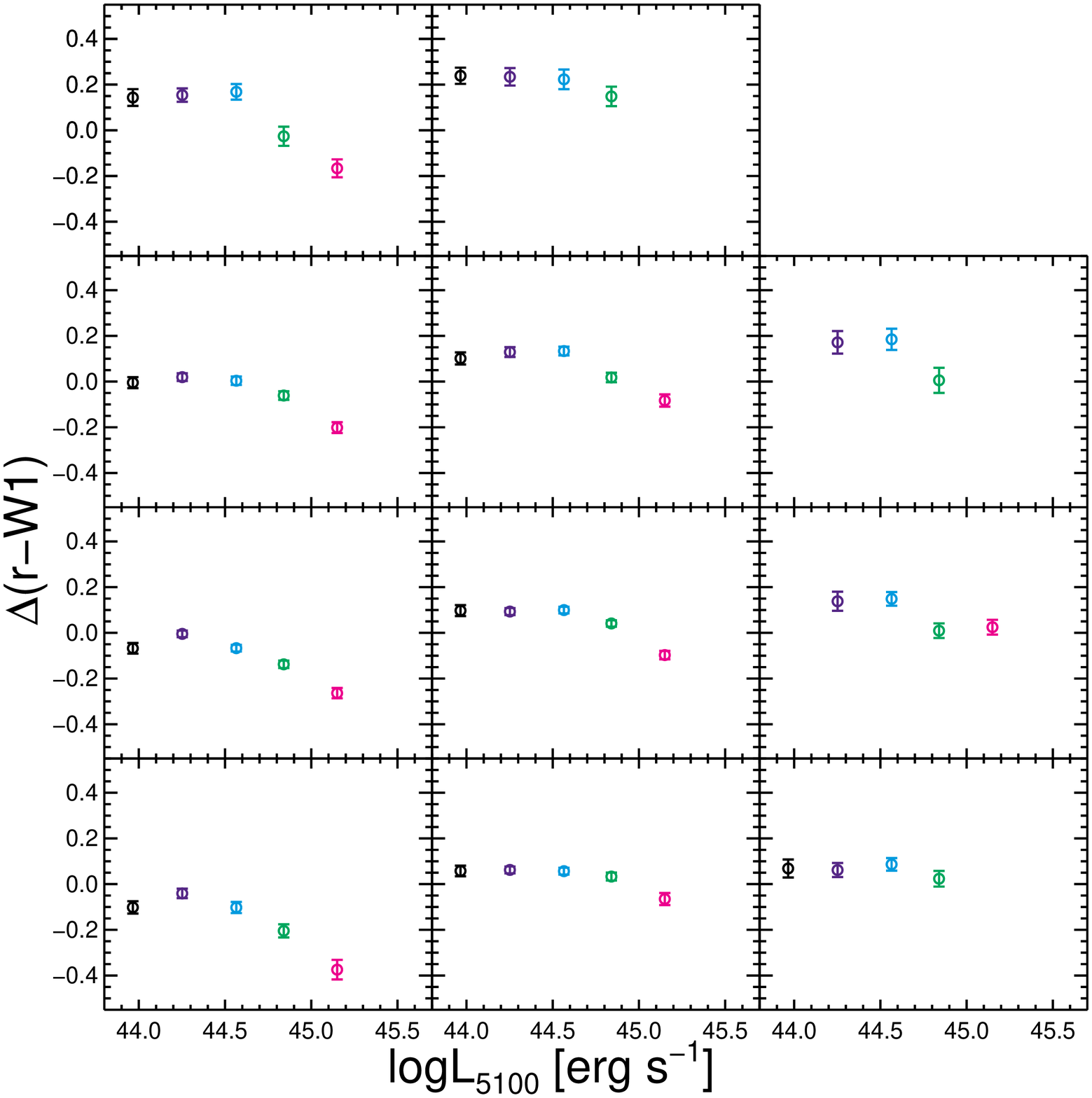}
 \caption{\textbf{A detailed look at the average optical-WISE color in the EV1 plane.} The same bins defined in Figure\ 1 are used. Error bars are the uncertainty in the mean, estimated by the standard deviation divided by the square root of the number of objects in the bin. At fixed $\Rfe$, we see increasing relative torus emission when FWHM$_\textrm{\hbeta}$ increases. This is consistent with the orientation scenario: larger FWHMs indicate more edge-on systems, which suffer more from geometric reduction (the $\cos I$ factor) and/or dust extinction in the optical than in the infrared. 
 }
 \label{fig:wise_more}
\end{figure*}

\begin{figure*}
\renewcommand\thefigure{E7}
 \centering
 \includegraphics[width=0.48\textwidth]{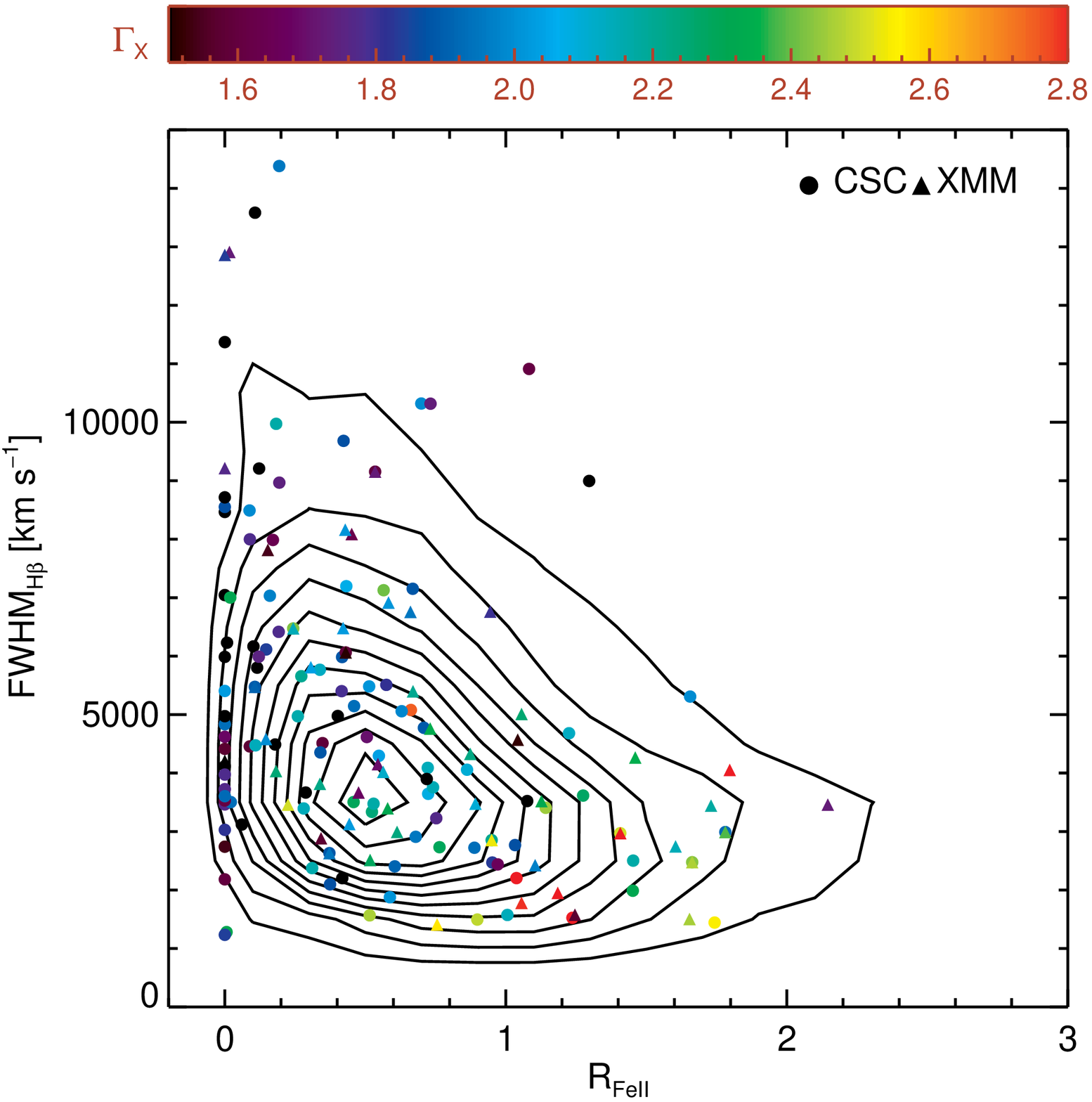}
 \caption{\textbf{Distribution in the EV1 plane in terms of X-ray properties.} The subset of our SDSS quasars with available measurements of their soft X-ray photon index $\Gamma_{\rm X}$ are shown. $\Gamma_{\rm X}$ increases (becomes softer) with increasing $\Rfe$, consistent with earlier findings\cite{Wang_etal_1996,Laor_etal_1997}. 
 }
 \label{fig:Xgamma}
\end{figure*}

\begin{figure*}
\renewcommand\thefigure{E8}
 \centering
 \includegraphics[width=0.48\textwidth]{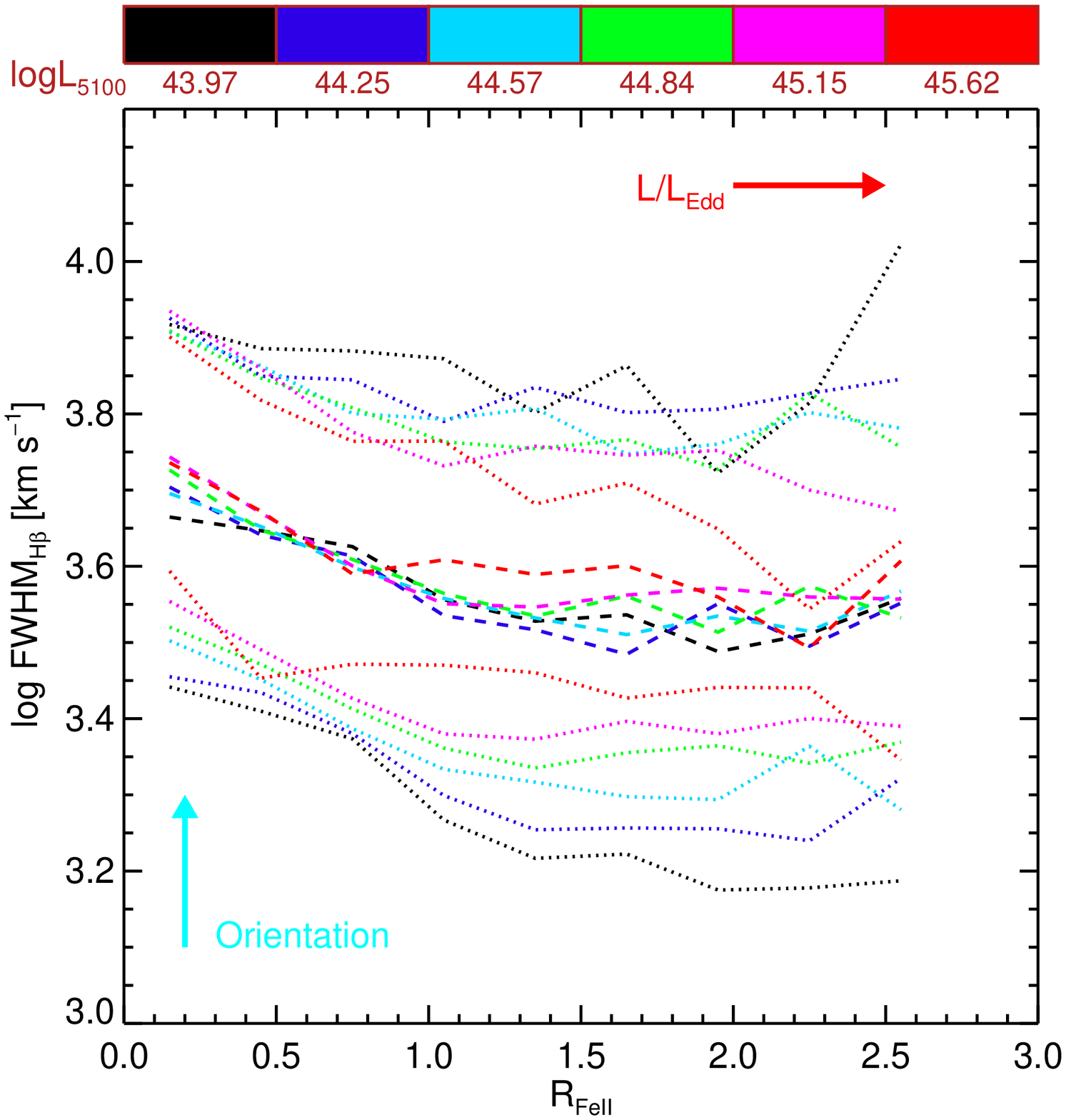}
 \caption{\textbf{The same EV1 plane as in Figure\ 1 in logarithmic FWHM$_\textrm{\hbeta}$}. The dashed lines show the running medium value as a function of $\Rfe$ and the dotted lines show the 16\% and 84\% percentiles, for objects in different luminosity bins. The distribution of FWHM$_\textrm{\hbeta}$ at fixed $\Rfe$ roughly follows a log-normal distribution, with a dispersion of $\sim 0.15-0.25$ dex, which we argued mostly comes from orientation-induced variations. Lower-luminosity objects tend to have slightly larger dispersion in FWHM$_\textrm{\hbeta}$, possibly cased by a broader Eddington ratio distribution at lower-luminosities, which introduces additional dispersion in FWHM$_\textrm{\hbeta}$. 
 }
 \label{fig:logFWHM}
\end{figure*}

\begin{figure*}
\renewcommand\thefigure{E9}
 \centering
 \includegraphics[width=0.48\textwidth]{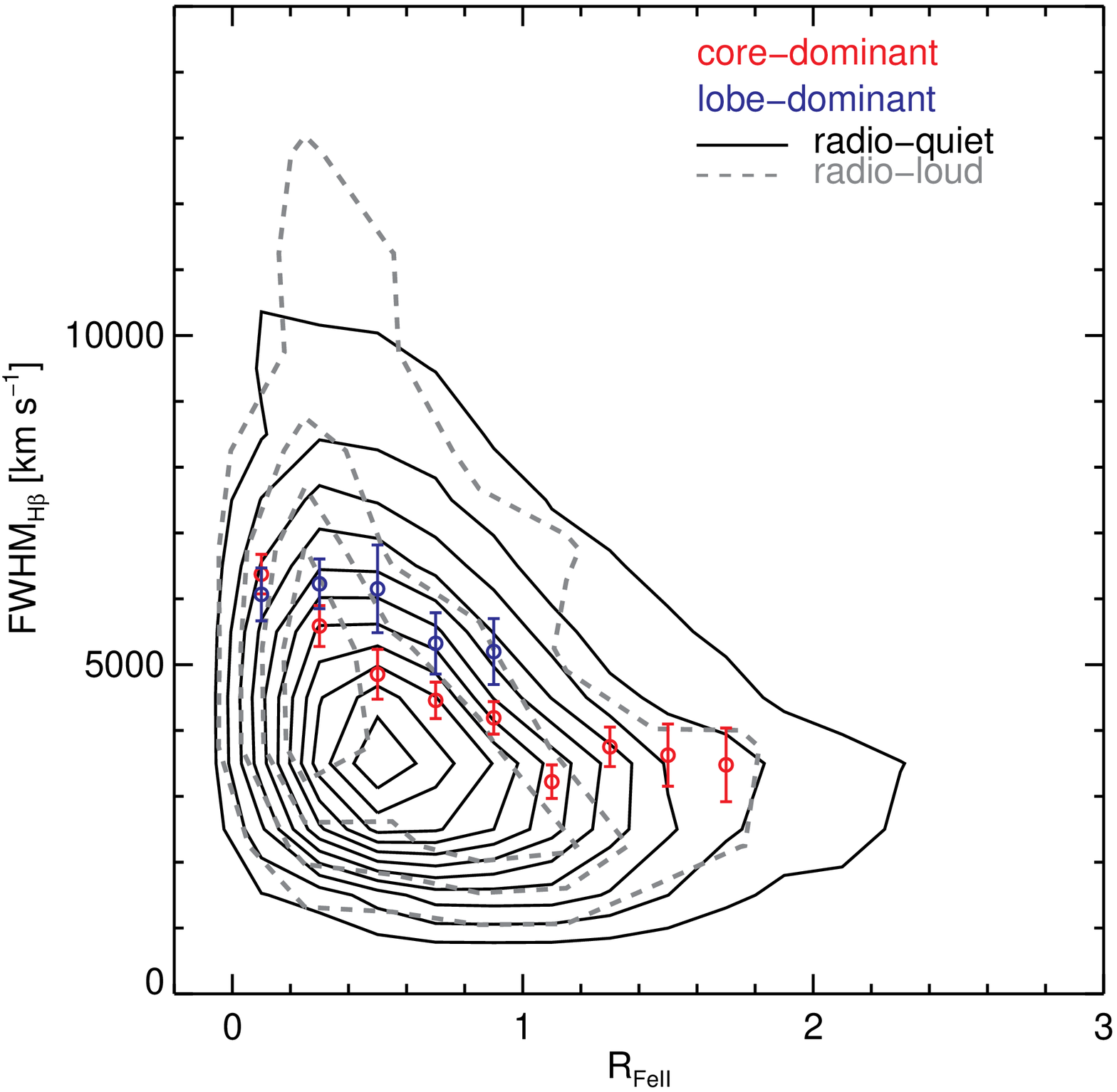}
 \caption{\textbf{Distributions of radio-loud and radio-quiet quasars in EV1 plane.} The radio-loud population shifts to lower $\Rfe$ and larger FWHM$_\textrm{\hbeta}$, compared with the radio-quiet population. We further divide the radio-loud quasars into core-dominant and lobe-dominant subsets, but we caution that our morphological classification is very crude, and there is potentially a large mixture of true morphological types between the two subsamples. The core-dominant (more pole-on) radio quasars have systematically smaller FWHM$_\textrm{\hbeta}$ compared with the lobe-dominant radio quasars, consistent with the hypothesis that orientation leads to variations in FWHM$_\textrm{\hbeta}$. The points with error bars are the median and uncertainty in the median in each $\Rfe$ bin.
 }
 \label{fig:radio}
\end{figure*}

\begin{figure*}
\renewcommand\thefigure{E10}
 \centering
 \includegraphics[width=0.8\textwidth]{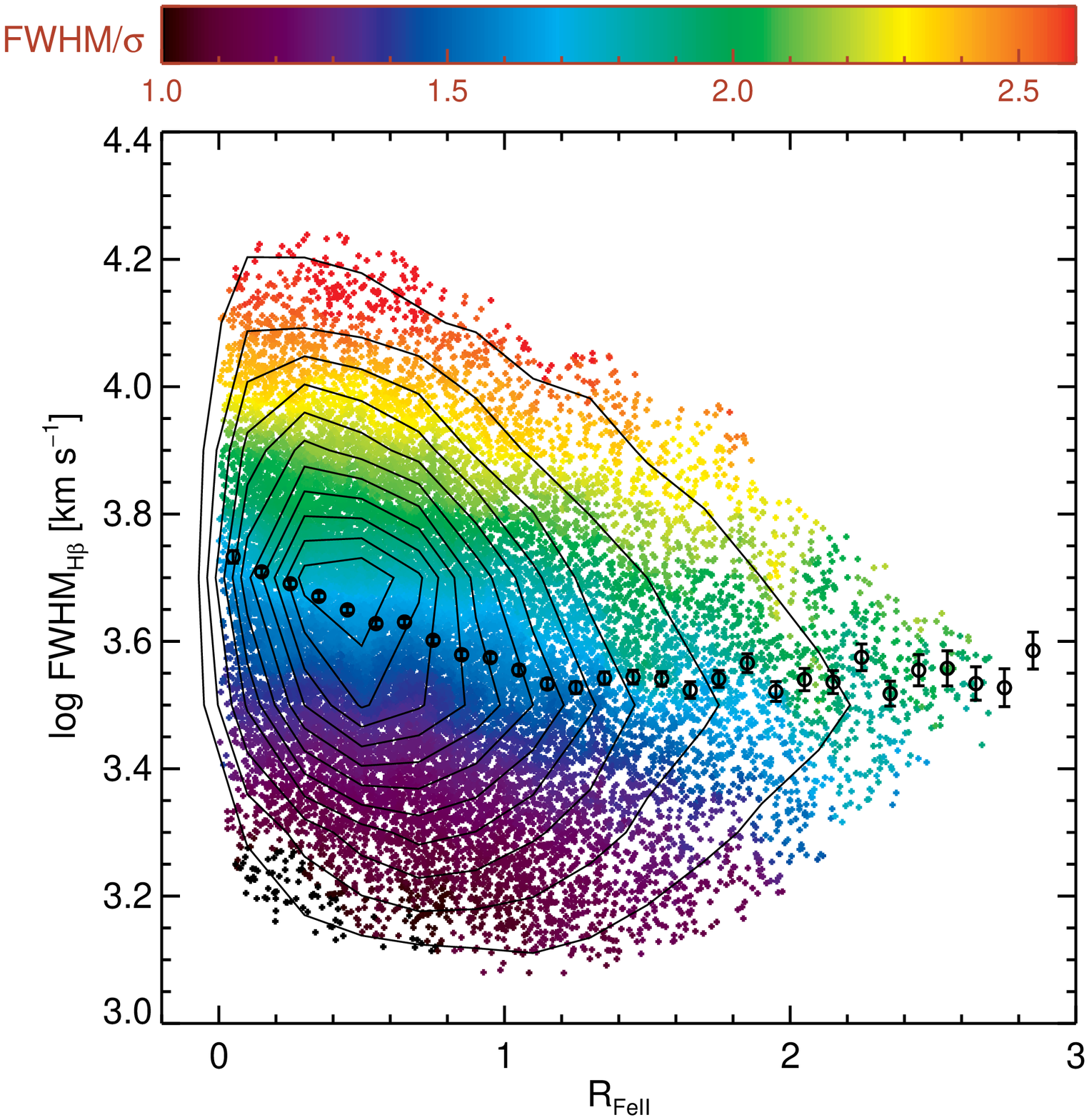}
 \caption{\textbf{Distribution in the EV1 plane color-coded by the FWHM$/\sigma$ ratio.} The distribution has been smoothed over a box of $\Delta\Rfe=0.2$ and $\Delta \log{\rm FWHM}_{\textrm{\hbeta}}=0.2$. We only show points for which there are more than 50 objects in the smoothing box to average. The black open circles show the median FWHM$_\textrm{\hbeta}$ at fixed $\Rfe$ (using all objects in that bin), with the error bars indicating the uncertainty in the median. The transition in FWHM$/\sigma$ reflects the change in orientation of the broad-line region disk relative to the line-of-sight.
 }
 \label{fig:fwhm_sigma_ratio}
\end{figure*}

\noindent \textbf{\huge Supplementary Information}

\section{Main Quasar Sample and Spectral Decomposition}\label{sec:sample}

The main quasar sample of this study is drawn from the SDSS DR7 quasar catalog (6), which contains more than 20,000 quasars at $z<0.9$ for which the \hbeta-\OIII\ region is covered in the SDSS spectra. We use the improved redshift estimates (28) for these SDSS quasars. Our SDSS quasar sample probes a redshift range $0.1\le z\le 0.9$ and a luminosity range of $L_{\rm 5100}\approx 10^{44-45.5}\ {\rm erg\,s^{-1}}$, where we adopt a flat $\Lambda$CDM cosmology with $\Omega_0=0.3$ and $H_0=70\,{\rm km\,s^{-1}Mpc^{-1}}$ throughout.

We measure the continuum and emission-line properties for \FeII, \hbeta\ and \OIII\ using functional fits to the spectra. The details of the spectral fits are described in early work (7, 29). In short, several continuum windows around the spectral region of \hbeta-\OIII\ is fit by a power-law continuum plus a \FeII\ template (1) to form a pseudo-continuum; this pseudo-continuum is subtracted from the spectrum, leaving an emission-line-only spectrum. A set of Gaussians is then fit to the broad and narrow emission lines. The narrow-line component of \hbeta\ is modelled using constraints from the narrow \OIII\ lines; we fix the offset and line width of narrow \hbeta\ to those of the narrow \OIIIab\ doublet. In cases where the \OIII\ doublet shows a prominent blueshifted wing, we fit \OIIIa\ and \OIIIb\ with double Gaussians, one for the core component and the other for the wing component. Narrow \hbeta\ is then tied to the core \OIII\ component. The broad \hbeta\ component is fit with three Gaussians. In addition, we fit two Gaussians to account for the narrow and broad \HeII\, $\lambda$4686. We determine the line FWHM and rest-frame equivalent width (EW) for broad \hbeta\ and \OIIIb, as well as the rest-frame EW of the optical \FeII\ emission within 4434-4684\ \AA. As common practice, we use the ratio, $R_{\textrm{\FeII}}={\rm EW_{FeII}/EW_{H\beta}}$, to indicate the strength of the optical \FeII\ emission. For other emission lines, the term ``strength'' refers to the rest-frame EW (or line luminosity normalized by the continuum luminosity), unless otherwise specified.

The spectral decomposition generates the emission-line-only spectra (i.e., with the pseudo-continuum removed) and the narrow-line-only spectra (i.e., with both the pseudo-continuum and the broad \hbeta\ component removed), with which we create stacked spectra binned in quasar properties.

\section{Summary of Quasar Properties Along Eigenvector 1 (EV1)}\label{sec:eve}

Here we summarize all the quasar properties that correlate with the EV1, using our large SDSS quasar sample, combined with other multi-wavelength datasets and additional quasar samples. This exercise updates the known EV1 correlations with improved samples and adds new quantities to the EV1 sequence. \\

\noindent \textbf{\OIII\ as an EV1 correlator}

The strong anti-correlation between the \OIII\ strength and the optical \FeII\ strength in the original EV1 correlation is shown in Figure\ 1. The dots are individual objects, color-coded by the average strength of the \OIIIb\ line (i.e., EW$_{\rm [OIII]}$); for each object, we compute an average EW$_{\rm [OIII]}$ by taking the median value of all the points within a smoothing box of $\Delta \Rfe=0.2$ and $\Delta {\rm FWHM_{H\beta}}=1000\ {\rm km\,s^{-1}}$ centered on that object. We plot the object only if there are more than 50 objects in the smoothing box. This is similar to measuring the average value from the composite spectrum of all the objects in that bin. This way we can better visualize the smooth trend in the average properties. Given the large number of objects in each smoothing box, the statistical uncertainty in the average value is negligible, with a median error of $\sim 0.016$ dex and $99.6\%$ of the time below $0.06$ dex.

The dominant trend in the colormap of Figure\ 1 is the decrease of the \OIII\ strength when $\Rfe$ increases, well known as the EV1 correlation (1). However, there is little vertical trend in the \OIII\ strength when FWHM$_\textrm{\hbeta}$ changes. We have broken down the distribution in individual luminosity bins, and found similar results; quasars with different luminosities occupy similar space in this plane, and follow the same trend with $\Rfe$ in terms of \OIII\ strength (although the average \OIII\ strength depends on luminosity, as described below). 

To investigate the behavior of the \OIII\ line in the $\Rfe$-FWHM$_\textrm{\hbeta}$ plane, we generate composite spectra in bins of $\Rfe$ and FWHM$_\textrm{\hbeta}$, and as a function of $L_{5100}$. The $\Rfe$-FWHM$_\textrm{\hbeta}$ bins are shown in the gray grid in Figure\ 1. In each bin, we divide the objects into 6 different luminosity bins, and generate the median composite spectra using the individual narrow-line-only (i.e., pseudo-continuum and broad \hbeta\ removed) spectra following our spectral decomposition. In detail, we take each narrow-line-only flux density spectrum, shift it to rest-frame, rebin onto a common wavelength grid (with a 1\,\AA\ dispersion) and then convert to a luminosity density spectrum using the luminosity distance of the quasar. The composite spectrum is generated by taking the median luminosity density at each wavelength pixel for all quasars in that $\Rfe$-FWHM$_\textrm{\hbeta}$-luminosity bin. An error spectrum is generated by dividing the 68\% semi-interquantile range of the luminosity densities by the square root of the number of objects contributing to each pixel. A composite spectrum is shown only if there are more than 60 objects in the stack for that bin. Typically there are more than 100 objects in each bin, and the statistical uncertainty (i.e., the error spectrum) of the resulting composite spectrum is by all means negligible compared to the observed trends (for example, median S/N$>5$ or much higher per 1\AA\ pixel within $\pm 1000\,{\rm km\,s^{-1}}$ of \OIII), and hence is suppressed in Fig.\ 2 and Extended Data Fig.\ E3 for clarity. Finally we scale each line luminosity density by the median continuum luminosity $L_{5100}$ of each $\Rfe$-FWHM$_\textrm{\hbeta}$-luminosity bin, such that the relative flux between different composites reflects the \OIII\ strength. In the last step we also corrected the average host contamination at 5100\,\AA, using the empirical relation derived using the same SDSS quasar sample (7), which only affects the lowest-luminosity bins with $\log (L_{\rm 5100,total}/{\rm erg\,s^{-1}})< 45.053$:
\begin{equation}
L_{\rm 5100,host}/ L_{\rm 5100,AGN} = 0.8052 - 1.5502x + 0.9121x^2 - 0.1577x^3
\end{equation}
where $x + 44 \equiv \log(L_{\rm 5100,total} /{\rm erg\,s^{-1}} ) < 45.053$.

Figure\ 2 shows that \OIII\ is weaker in larger $\Rfe$ bins but similar in different FWHM$_\textrm{\hbeta}$ bins when $\Rfe$ is fixed, as indicated in Figure\ 1. In addition, \OIII\ is weaker for higher quasar luminosities, known as the \OIII\ Baldwin effect (8-10). However, as already mentioned above, at fixed luminosity, the trend of \OIII\ strength with $\Rfe$ still holds from the left to the right columns in Figure\ 2. This means that the change (i.e., the EV1 anti-correlation between \FeII\ and \OIII\ strength) is either due to a systematic change in BH mass (hence Eddington ratio), or, less likely, changes in the narrow-line region properties as $\Rfe$ increases.  

Another notable feature in Figure\ 2 is that the \OIII\ line profile can be decomposed into a core component, centered nearly at the systemic velocity, and a wing component, blueshifted from the systemic velocity by a few hundred ${\rm km\,s^{-1}}$, similar to those seen in individual objects. For the individual composite spectra shown in Figure\ 2, we fit the two components with double Gaussians for the \OIIIab\ doublet and obtain the total flux, FWHM, and velocity shift of each component. We show the resulting \OIIIb\ luminosity, normalized by $L_{5100}$, as a function of $L_{5100}$ in Extended Data Figure E1, for both the core and the wing components. Here again $L_{5100}$ is corrected for host contamination following the above empirical recipe, and different colors show different $\Rfe$ bins. The measurement uncertainties of \OIII\ properties are estimated using Monte Carlo trials of mock spectra, generated using the estimated error arrays in the coadded spectra. These measurement uncertainties are generally negligible compared to the observed trends. The weakening in the total \OIII\ emission as luminosity increases is primarily due to the weakening of the core component -- the wing component remains more or less constant. On the other hand, when $\Rfe$ increases, both \OIII\ components are weakened, with the wing component again weakening at a slower pace.  

{Extended Data Figure\ E2} shows the detailed kinematic properties of \OIII.  The most significant trends are the slight correlation with luminosity of the FWHM of the core component and the velocity offset from systemic velocity of the wing component.  In addition, there is a slight correlation between the velocity offset of the wing component and \FeII\ strength. 

The qualitative and quantitative differences between the core and wing \OIII\ components suggest that the two may have different origins. While the core component may be dominantly excited by photoionization, the blueshifted wing component may be more related to shocks associated with outflows (11). As luminosity or Eddington ratio increases, the underlying spectral energy distribution (SED) from the accreting BH changes shape (30,31), which leads to changes in the core \OIII\ strength (see more discussions in \S\ref{sec:main}). Of course, these are simply speculations and detailed follow-up studies are needed to test various possibilities. 

A point worth noting here is that the peak of the core \OIII\ component is generally consistent with the systemic velocity (28) to within $\sim 50\,{\rm km\,s^{-1}}$, and is also consistent with the peak from low-ionization narrow lines such as \OII\ (see Extended Data Fig.\ E3). However, a single Gaussian fit to the overall \OIII\ profile may yield a blueshift relative to systemic, due to the contamination from the wing component. \\

\noindent \textbf{Other Emission Lines}

We find that all the high-ionization (e.g., with an ionization potential equal to or higher than that of \OIII, $\sim 35$ eV) lines covered in the SDSS spectrum have similar behaviors as \OIII\ in terms of the change of line strength as $\Rfe$ and luminosity change. Low-ionization lines (such as \MgII, \OII, \SII), on the other hand, have similar but slightly weaker trends as $\Rfe$ or luminosity changes. We show in {Extended Data Figure\ E3} several examples of \hbeta, \MgII, \OII\ and \NeV, in the same bins defined in the $\Rfe-$FWHM$_\textrm{\hbeta}$ plane. As a high-ionization line, \NeV\ also shows a blueshifted wing component similar to \OIII. Therefore, the original EV1 relation between \OIII\ and optical \FeII\ strengths applies to all high-ionization lines. 

There are two additional results from {Extended Data Figure\ E3}: (a) the UV \FeII\ strength near the \MgII\ line does not correlate with the optical \FeII\ strength; (b) the broad \MgII\ FWHM is correlated with FWHM$_\textrm{\hbeta}$, as confirmed in individual quasars (29, 32).

We augment these optical measurements with \CIV\ properties, using an additional sample of $\sim 130$ low-redshift quasars with both optical spectra covering \hbeta\ (24) and UV spectra covering \CIV\ from various data archives of space-based satellites (FUSE, HST, IUE). We measure the properties of \CIV, \hbeta\ and \FeII\ in the same way as described above. {Extended Data Figure\ E4} shows the distribution of this additional sample in the EV1 plane, color-coded by the \CIV\ EW. A clear trend of decreasing \CIV\ strength with $\Rfe$ is observed, similar to other high-ionization lines. This trend is real, since the typical measurement uncertainty in \CIV\ EW is only $\sim 7\%$ (relative). Although we do not have enough statistics to divide this low-redshift sample by luminosity, \CIV\ strength is well known to decrease with luminosity based on studies of high-redshift quasars, discovered as the original Baldwin effect (10). These trends also seem to correlate with the kinematic properties of \CIV\ such as the blueshift and line asymmetry (2, 33). However, we note that the geometry of the \CIV\ broad-line region (BLR) may be different from that for the \hbeta\ BLR --- the main focus of this work --- as inferred from the different properties of these broad lines (22, 32, 34, 35). \\

\noindent \textbf{Mid-Infrared Properties}

{Extended Data Figure\ E5} examines trends with optical-infrared color.  We have matched our SDSS quasar sample to the WISE all-sky survey source catalog (12). More than 80\% of our quasars have a WISE counterpart detected within 0.5'' of the optical position, the typical astrometric precision for sources brighter than 16 mag in the WISE $W1$ band ($3.4\,\mu{\rm m}$). In the following we only use the $W1$ band, but we emphasize that similar results are found for the other WISE bands at longer wavelengths as well. The top panel shows the distribution of quasars in the EV1 plane, colored-coded by the $r-W1$ color, where the optical SDSS $r$-band magnitudes have been corrected for Galactic extinction. To limit the effect of band shifting due to redshift, we restrict our sample to $0.4<z<0.8$. To first order, there is a trend of redder optical-WISE color toward higher $\Rfe$, indicating higher relative flux in the mid-infrared than in the optical when $\Rfe$ increases. The mid-infrared emission in WISE bands mostly comes from hot dust (here we use ``hot dust'' to refer to quasar-heated torus dust, as opposed to the hottest dust at the inner edge of the torus) in the torus region, while the optical continuum comes from the accretion disk (AD). Therefore this trend suggests more efficient torus emission in high-$\Rfe$ objects, where the Eddington ratio is higher as argued in this work. Of course, the actual structure and emission of the torus may be substantially more complicated, which may explain the clumpy features in Extended Data Fig.\ E5 for the optical-IR colors. Our analysis is not intended to fit realistic torus models, but simply to show that, to first order, there is a clear enhancement of torus emission along the EV1 sequence.

We can also study the distribution of the excess $r-W1$ color, $\Delta(r-W1)$, the deviation from the median color at each redshift. This color excess removes the effect of band shifting with redshift and can be used to indicate the relative strength of torus to AD emission. Such a colormap is shown in the right panel of {Extended Data Figure\ E5}, using all SDSS quasars with the relevant measurements. A similar global trend is observed along the EV1 sequence, with higher $\Rfe$ quasars having more torus emission relative to the disk emission. 

It might be tempting to connect the anti-correlation between the \OIII\ strength and $\Rfe$ to the correlation between the torus emission and $\Rfe$. If the increasing (relative) infrared emission along EV1 is caused by an increasing torus covering factor, the resulting solid angle of the \OIII\ ionization cone will decrease along EV1, which will lead to a decreasing \OIII\ strength. However, the relative torus emission only increases by $\sim 40\%$ from $\Rfe=0$ to $\Rfe=2$, while the average \OIII\ strength decreases by at least a factor of 5 over the same $\Rfe$ range (Figure\ 1). Therefore, a pure geometric effect (enhanced torus covering factor along EV1) cannot explain the observed EV1 correlation in terms of the \OIII\ strength --- the change in the underlying quasar SED along EV1 is a more plausible explanation (see \S\ref{sec:main}).  

In a similar spirit to our investigations on the emission lines, we study the average optical-infrared color in bins of FWHM$_\textrm{\hbeta}$ and $\Rfe$, and as a function of luminosity. {Extended Data Figure\ E6} shows the results, in terms of the average color excess $\Delta(r-W1)$ in each bin. We see increased relative torus emission when $\Rfe$ increases, as in {Extended Data Figure\ E5}. In fixed $\Rfe$ bins, we see increasing relative torus emission when FWHM$_\textrm{\hbeta}$ increases. This effect is subtle, only $\sim 20-30\%$ in the relative sense, which may explain why this was not detected in small samples with substantial object-by-object variations (36). While alternative explanations may exist, this is consistent with the orientation scenario, wherein larger FWHMs indicate more edge-on systems, where the optical luminosity is reduced by a $\cos I$ factor (where $I$ is the inclination angle with $I=90^\circ$ corresponding to edge-on), and/or suffer more from dust extinction than the infrared luminosity. To test the latter possibility, we have checked the optical $g-i$ colors in the same bins as in Extended Data Fig.\ E6, but did not find strong evidence that more edge-on systems suffer more extinction in the optical (after accounting for the apparent reddening in the optical due to host contamination in low-luminosity bins). We therefore conclude, under the orientation scenario, that the geometrical reduction in the optical luminosity from an inclined AD is the main reason for the increasing relative torus emission when FWHM$_\textrm{\hbeta}$ increases (see more discussion in \S\ref{sec:main}). Finally, there is a mild trend of decreasing relative torus emission when luminosity increases (especially at the highest luminosities), in accord with earlier studies (37). 

The finding of an EV1 sequence in terms of mid-infrared emission is new, and motivates more realistic torus models to constrain quantitatively the structure of the torus along the quasar EV1 sequence. \\

\noindent \textbf{X-ray Properties}

Finally, it is known that the soft X-ray photon index $\Gamma_{\rm X}$ ($f_E\propto E^{-\Gamma_{\rm X}}$) increases along the EV1 sequence (3, 5). We matched our SDSS quasars with the Chandra Source Catalog (CSC, 38) and the XMM-Newton Serendipitous Catalog (39) with a matching radius of 1'', and found $\sim 120$ matches. We show the distribution of these matched quasars in the EV1 plane in {Extended Data Figure\ E7}, color-coded by $\Gamma_{\rm X}$, which for Chandra sources were taken directly from the CSC master table (which are consistent with those measured in [40]) and for XMM-Newton sources were taken from [41]. The measurement uncertainties in $\Gamma_{\rm X}$ are typically $\sim 10\%$ (relative uncertainty; 1$\sigma$) for the Chandra sources, and $\sim 20\%$ for the XMM sources. We find that $\Gamma_{\rm X}$ increases systematically with $\Rfe$,  consistent with early results (3, 5) and with the known tendency for $\Gamma_{\rm X}$ to steepen with increasing Eddington ratio (5, 42). \

\section{Orientation-Induced Dispersion in FWHM$_\textrm{\hbeta}$}\label{sec:orient}

It is known from observations of radio-loud quasars that FWHM$_\textrm{\hbeta}$ can be affected by orientation (21, 22), such that high-inclination (more edge-on) systems have, on average, larger FWHM$_\textrm{\hbeta}$.  The \hbeta-emitting portion of the BLR presumably has a flattened geometry aligned with the AD, whose inclination can be estimated from the radio morphology of the jet. Although most ($\sim 90\%$) quasars are radio-quiet, it is generally expected that orientation should play some role in determining the measured line-of-sight line widths of the general quasar population. 

Motivated by the lack of variation of \OIII\ properties on FWHM$_\textrm{\hbeta}$ at fixed $\Rfe$, as well as observations of radio-loud quasars, we have suggested that most of the dispersion in FWHM$_\textrm{\hbeta}$ seen at fixed $\Rfe$ is actually due to the orientation effect of a flattened BLR geometry. We now examine this possibility in detail. \\

\noindent\textbf{Dispersion in FWHM$_\textrm{\hbeta}$} 

We show in {Extended Data Figure\ E8} the dispersion of the logarithmic FWHM$_\textrm{\hbeta}$ as a function of $\Rfe$ and in different luminosity bins. The dispersion in FWHM$_\textrm{\hbeta}$ is roughly $0.15-0.25$ dex at fixed $\Rfe$. There is a slight trend of increasing dispersion in FWHM$_\textrm{\hbeta}$ when luminosity decreases, possibly caused by the possibility that the Eddington ratio distribution (hence BH mass distribution) at fixed luminosity is broader at faint quasar luminosities. This additional dispersion in BH mass is translated into FWHM$_\textrm{\hbeta}$, and broadens the total dispersion in FWHM$_\textrm{\hbeta}$ at fixed luminosity. This additional broadening is nevertheless small, $\sim 0.1$ dex in FWHM$_\textrm{\hbeta}$ comparing the distributions at the lowest and highest luminosity bins. This is in line with the idea that most of the dispersion in FWHM$_\textrm{\hbeta}$ at fixed $\Rfe$ is due to orientation.

The average FWHM$_\textrm{\hbeta}$ decreases with $\Rfe$, which could be due to real changes in the virial velocity of the BLR gas, in which case the average FWHM$_\textrm{\hbeta}$ can be used to estimate the BH mass along the $\Rfe$ sequence.  On the other hand, the dispersion in FWHM$_\textrm{\hbeta}$ at fixed $\Rfe$ is due largely to an orientation effect as argued in this work, implying a flattened geometry for the BLR. In the latter case, the variation in FWHM$_\textrm{\hbeta}$ at fixed $\Rfe$ does not reflect changes in the BH mass. \\

\noindent\textbf{Dependence on Radio Morphology} 

As mentioned earlier, a small subset ($\sim 10\%$) of quasars are radio-loud, and the radio jet morphology can be used to estimate the orientation of the AD. Studies of resolved radio morphology of small (of order $50$) samples of low-redshift radio quasars have shown an anti-correlation between the radio core dominance and FWHM$_\textrm{\hbeta}$ (21, 22). Here we perform a similar test using our SDSS quasar sample. A small fraction of quasars in our sample are detected in the FIRST radio survey (43). For simplicity we define the FIRST-detected quasars as the ``radio-loud'' sample and the undetected quasars as the ``radio-quiet'' sample, which is different from the traditional definition of radio-loudness. As FIRST is a shallow radio survey, $\sim 90\%$ of the FIRST-detected quasars satisfy the traditional definition of being radio-loud (7). We can determine a rough radio morphology based on the number of FIRST sources detected around the quasar (44): for quasars that have only one FIRST source within 30'' we match them again to the FIRST catalog with a matching radius 5'' and classify the matched ones as core-dominant radio quasars. Those that have multiple FIRST source matches within 30'' are classified as lobe-dominated. There are $\sim 1600$ core-dominant and $\sim 400$ lobe-dominated quasars in our sample for which we also have \FeII\ and \hbeta\ measurements. Although this morphological classification of the radio-detected quasars is by no means perfect and introduces significant mixture of true radio morphologies between the two subsets, on average we expect the core-dominant objects are more pole-on than the lobe-dominant objects. There is no significant difference in the mean quasar luminosity between the two radio subsets. 

{Extended Data Figure\ E9} shows the contours in the $\Rfe$ and FWHM$_\textrm{\hbeta}$ plane for both the radio-loud and radio-quiet quasars in our SDSS sample. We also show the median FWHM$_\textrm{\hbeta}$ at fixed $\Rfe$, for both core-dominant and lobe-dominant quasars, in this EV1 plane. At fixed $\Rfe$, core-dominant objects have systematically lower FWHM$_\textrm{\hbeta}$ than lobe-dominant objects, consistent with the orientation hypothesis. In addition, the radio-detected population shifts to lower $\Rfe$ and larger FWHM$_\textrm{\hbeta}$ compared with the radio-quiet population. This is a well known result (2, 16, 33), consistent with the notion that radio-loud quasars preferentially reside in more massive and lower Eddington ratio systems (16, 45). \\

\noindent\textbf{Tests with Independent BH Mass Estimates} 

The orientation scenario posits that most of the dispersion in FWHM$_\textrm{\hbeta}$ at fixed $\Rfe$ does not reflect the changes in the virial velocity of the BLR. Therefore the traditional virial BH mass estimates based on FWHM$_\textrm{\hbeta}$ will lead to a FWHM-dependent bias. Below we test this idea using additional quasar samples for which the BH masses can be estimated using independent methods. 

The first sample are the 29 local AGNs with reverberation mapping (RM) data, for which stellar velocity dispersion measurements are available (23, 46-52). We have measured the \hbeta\ and \FeII\ properties using single-epoch spectra (24) for these objects, with the same fitting procedure as for the SDSS quasars. We estimate the BH masses using the observed relation between stellar velocity dispersion and the BH mass for local inactive galaxies (25). The virial coefficient, which determines the relation between FWHM$_\textrm{\hbeta}$ and the underlying virial velocity, can be defined as:
\begin{equation}
f=\frac{GM_{\rm BH,\sigma_*}}{R_{\rm BLR}{\rm FWHM^2_{\textrm{\hbeta}}}}\ ,
\end{equation}
where $M_{\rm BH,\sigma_*}$ is the BH mass estimated from the $M_{\rm BH}-\sigma_*$ relation (25), and $R_{\rm BLR}$ is the BLR size measured from RM for these objects (15, 46, 53-55). We neglect the time variability between the single-epoch spectroscopy and the RM measurements of the BLR size. If FWHM$_\textrm{\hbeta}$ traces the virial velocity well, then at fixed BH mass there should be little trend of $f$ with FWHM$_\textrm{\hbeta}$. The dominant uncertainty in the estimation of $f$ comes from the intrinsic scatter in the $M_{\rm BH}-\sigma_*$ relation (25) rather than measurement errors (i.e., $\sigma_{\log f}\sim 0.3$ dex). This systematic uncertainty in $f$ is substantial, which may be responsible for most of the vertical scatter in Figure\ 4, but cannot drive a correlation with FWHM$_\textrm{\hbeta}$.

Figure\ 4 shows the results, where we divide the $\sigma_*$-based BH masses into three bins. There is a clear segregation among objects with different BH masses: more massive BHs tend to have larger FWHM$_\textrm{\hbeta}$ on average, indicating that FWHM$_\textrm{\hbeta}$ does play some role in determining the BH mass. This statement is also supported by the fact that when FWHM$_\textrm{\hbeta}$ is used in the single-epoch virial mass estimates, the correlation between these virial masses with $\sigma_*$-based BH masses is closer to a linear relation than that without using FWHM$_\textrm{\hbeta}$ (i.e., assuming a constant FWHM for all objects). However, in each mass bin, there is an apparent trend of the virial coefficient with FWHM$_\textrm{\hbeta}$, suggesting that most of the variance in FWHM$_\textrm{\hbeta}$ at fixed BH mass is uncorrelated with the virial velocity (hence the BH mass). Naturally, orientation-induced variations in FWHM$_\textrm{\hbeta}$ explain the observed trend of $f$ with FWHM$_\textrm{\hbeta}$ at given BH mass. The dispersion in FWHM$_\textrm{\hbeta}$ at fixed true BH mass (based on $\sigma_*$) is about 0.2 dex, introducing a factor of $\sim 2.5$ scatter in the virial BH mass estimates based on FWHM$_\textrm{\hbeta}$ (from single-epoch spectroscopy) at fixed true BH mass. 

To improve the small-number statistics based on the RM sample, we use the stellar velocity dispersion measurements for a sample of $\sim 600$ low-redshift SDSS AGNs obtained from spectral decompositions of the AGN spectrum and the host spectrum (26). These SDSS AGNs typically have lower luminosity than our SDSS quasar sample in this study, such that the host galaxy spectrum can be decomposed from the AGN spectrum. We estimate the BLR sizes for this sample using the observed tight $R_{\rm BLR}-L$ relation based on RM (27), and the BH masses using the same $M_{\rm BH}-\sigma_*$ relation. The results are shown as colored dots in Figure\ 4. Although there are different systematics with the two samples, it is reassuring that both produce similar trends. \\

\noindent\textbf{Dependence on Broad Line Shape} 

If the \hbeta-emitting portion of the BLR has a flattened, disklike geometry, as we suggest, then FWHM$_\textrm{\hbeta}$ is sensitive to orientation, as it measures the core of the line profile. The BLR gas motion should also include a turbulent component, whose velocity distribution is more isotropic. The second moment of the line (i.e., line dispersion, $\sigma$) is more sensitive to this isotropic component than FWHM, as its measurement includes the extended wings of the line profile. Therefore, the shape of the broad line as measured by the ratio FWHM/$\sigma$ should be an indicator of the orientation of the disk component of the BLR (56, 57). 

{Extended Data Figure\ E10} shows the EV1 plane in which the objects are color-coded by FWHM$/\sigma$. We measure $\sigma$ of broad \hbeta\ from our multiple-Gaussian model fits. This approach allows us to measure $\sigma$ without the need to truncate the spectrum at some wavelength in order to avoid noisy wings. A pattern of FWHM$/\sigma$ is apparent, such that at fixed $\Rfe$ FWHM$/\sigma$ increases toward larger FWHM$_\textrm{\hbeta}$. This is expected in our orientation scenario: at fixed $\Rfe$, FWHM$_\textrm{\hbeta}$ increases when the BLR disk is viewed more edge-on, but $\sigma$ is less affected by orientation, leading to a larger FWHM$/\sigma$ ratio when the inclination increases. It is remarkable that the color transition in {Extended Data Figure\ E10} is not purely vertical, but more or less parallel to the average FWHM$_\textrm{\hbeta}$ at each $\Rfe$. This reinforces the idea that orientation governs the distribution of FWHM$_\textrm{\hbeta}$ at fixed $\Rfe$, and the average FWHM$_\textrm{\hbeta}$ at fixed $\Rfe$ corresponds to the average inclination angle of the BLR disk. 

\section{Details on the Clustering Analysis}\label{sec:clustering}

Our investigation so far strongly supports the idea that the vertical dispersion (in FWHM$_\textrm{\hbeta}$) in the EV1 sequence is largely an orientation effect. A corollary is that samples divided using the FWHM-based virial BH masses will have large overlap in their true masses. On the other hand, the optical \FeII\ strength $\Rfe$ is most likely correlated with the Eddington ratio, and hence for our main SDSS sample the average BH mass should decrease from left to right in the EV1 plane. Most of the quasars in our main SDSS sample are too luminous to decompose the spectrum to estimate the host galaxy stellar velocity dispersion to yield a BH mass estimate independently without using FWHM$_\textrm{\hbeta}$.

To circumvent this problem, we turn to quasar clustering to infer the sample-averaged BH mass. More massive BHs are associated with more massive galaxies, which are in turn more strongly clustered (18). This exercise has been challenging previously, because detecting a clustering difference between low-mass and high-mass quasars requires good sample statistics, and the largest quasar samples to date still do not have enough statistics to measure a potential clustering difference among different quasar subsamples using auto-correlation functions (58).

The recent large spectroscopic surveys of SDSS-III (59) provide the largest spectroscopic sample of massive galaxies (19) at $0.3<z<0.9$, making it possible to cross-correlate quasars with galaxies (20) in the same redshift range for which we have the relevant quasar EV1 spectral measurements. The usage of a much larger galaxy sample to cross-correlate with the quasar sample can greatly improve quasar clustering measurements by reducing shot noise from low pair counts in quasar auto-correlation function measurements. The clustering difference we look for is subtle and requires superior measurement quality. Cross-correlation boosts the signal-to-noise ratio in the clustering measurements by a factor of $\sim\sqrt{N_{G}/N_{Q}}$ over the auto-correlation of quasars, where $N_G$ and $N_Q$ are the number of galaxies and quasars in our cross-correlation samples, respectively. The cross-correlation function (CCF) is determined by the auto-correlation functions of both sets of tracers; for the same galaxy sample, a stronger cross-correlation signal with a quasar subset will indicate a stronger intrinsic clustering of this quasar subset.
Our cross-correlation analysis represents the best measurement of quasar clustering in this redshift range, where our current SDSS quasar sample lies. We follow the same approach as Shen et al.\ (20) to measure the CCF between different quasar subsets and the galaxy sample, to see if a difference can be detected. The details regarding the galaxy sample, the cross-correlation technique and error estimation can be found in that paper. Here we only give a brief overview of the technical details regarding the clustering (cross-correlation) measurements. 

We select the SDSS DR7 quasars (6) and the SDSS-III DR10 CMASS galaxies (19) that are in the same overlapping area ($\sim 4100\,{\rm deg}^2$). We remove a relatively small number of quasars that were not targeted by the final quasar target selection algorithm (60) in SDSS DR7 to construct a flux-limited ($i<19.1$) quasar sample. Both the CMASS galaxy and the quasar samples are restricted to $0.3<z<0.9$, where most of the CMASS galaxies lie. Most quasars in our clustering sample then have the proper spectral measurements in the \hbeta-\OIII\ region. Our final clustering samples include $\sim 350,000$ galaxies and $\sim 7,800$ quasars. Random catalogs were generated using the same angular geometry and redshift distributions of the CMASS galaxy sample used. 

Following the general practice of clustering measurements, we estimate the 1D and 2D redshift space correlation functions $\xi_s(s)$ and
$\xi_s(r_p,\pi)$ using the Davis \& Peebles estimator (61):
$QG/QR -1$, where $QG$ and $QR$ are the normalized numbers of quasar-galaxy
and quasar-random pairs in each scale bin, $s$ is the pair separation in
redshift space, and $r_p$ ($\pi$) is the transverse (radial) separation in
redshift space.  To reduce the effects of redshift distortions, we use the projected correlation function
 (61)
\begin{equation}\label{eqn:DP}
  w_p(r_p) = 2\int_0^\infty d\pi\ \xi_s(r_p,\pi)\ .
\end{equation}
In practice we integrate $\xi_s(r_p,\pi)$ to $\pi_{\rm max}=70\,h^{-1}$Mpc,
and this cutoff is taken into account when fitting a model correlation function to the data. 

To estimate errors in the projected correlation function $w_p(r_p)$, we use jackknife resampling by dividing the clustering samples into
$N_{\rm jack}$ spatially contiguous and roughly rectangular regions with equal area, and creating
$N_{\rm jack}$ jackknife samples by excluding each of these regions in turn. We measure the
correlation function for each of these jackknife samples, and the covariance
error matrix is estimated as:
\begin{equation}
{\rm Cov}(i,j)=\frac{N_{\rm jack}-1}{N_{\rm jack}}\sum_{l=1}^{N_{\rm jack}}(\xi_i^l-\bar{\xi}_i)(\xi_j^l-\bar{\xi}_j)\ ,
\end{equation}
where indices $i$ and $j$ run over all bins in the correlation function, and
$\bar{\xi}$ is the mean value of the statistic $\xi$ over the jackknife
samples. The full covariance matrices will be used in our model fitting to the data. 

We consider two divisions of our main quasar sample, aiming at separating high-mass quasars from low-mass ones. The first division is based on the FWHM-based virial BH masses (7, 17); we divide the quasar sample by the median virial BH mass. The second division is based on $\Rfe$, whereby we divide the sample by the median $\Rfe$ value. We then measure the CCF between the two subsamples and the galaxy sample, and compare the measurements for both divisions. In both cases the redshift evolution of clustering is negligible given the narrow redshift range of our samples and the similar redshift distributions among different quasar subsamples.

The resulting CCFs are shown in Figure\ 3 for the two sample divisions. To compare the clustering strength among different samples, we fit simple power-law models, $\xi(r)=(r/r_0)^{-\gamma}$, to the CCFs. We fit to the range $r_p=0.2-50\ h^{-1}$ Mpc and use the full covariance matrix in the $\chi^2$ fit (20). We fix $\gamma=1.7$, the best-fit slope for the whole sample, and use the best-fit correlation length $r_0$ to indicate the clustering strength. For the division based on virial BH masses, we do not detect a significant difference ($1.64\sigma$) in the clustering strength, with $r_0=6.31\pm0.25\ h^{-1}$ Mpc and $6.95\pm0.30\ h^{-1}$ Mpc for the two quasar subsamples. This is expected: as we argued before, orientation leads to a large dispersion in FWHM$_\textrm{\hbeta}$ (and hence virial BH masses) and dilutes the intrinsic difference in BH masses between the two subsamples. On the other hand, we detect a significant ($3.48\sigma$) difference in the clustering strength when the quasar sample is divided by $\Rfe$, with $r_0=5.80\pm0.29\ h^{-1}$ Mpc and $7.11\pm0.24\ h^{-1}$ Mpc for the two quasar subsamples. This result strengthens our earlier point that the sequence from left to right in the EV1 plane is increasing in Eddington ratio, and hence on average decreasing in BH mass for our main quasar sample. This is by far the only significant detection to date of the dependence of quasar clustering on a physical property.

\section{The Quasar Main Sequence}\label{sec:main}

The observed systematic trends of multi-wavelength quasar properties in the EV1 plane ($\Rfe$ versus FWHM$_\textrm{\hbeta}$) lead to a simple, coherent picture of broad-line quasars: (a) $\Rfe$ is primarily correlated with Eddington ratio, although there might be a significant scatter around this correlation, such that at fixed $\Rfe$ there is still a dispersion of Eddington ratios; (b) the range of FWHM$_\textrm{\hbeta}$ at fixed $\Rfe$ includes a substantial component due to orientation effects such that more edge-on systems have on average larger FWHM$_\textrm{\hbeta}$, indicating a flattened BLR geometry, although the current analysis does not constrain the detailed structure of the postulated BLR disk, such as its thickness and radial extent. 

Under this framework, all observed correlations between physical properties and $\Rfe$ can be interpreted as due primarily to changes in the Eddington ratio. These physical properties include: strength of narrow emission lines (in particular the high-ionization lines), relative amount of torus emission, and X-ray spectral slope. All these properties are relevant to processes in the proximity of the accreting BH, and thus are likely tied to the accretion process itself. Changes in the Eddington ratio will lead to changes in the underlying SED from optical to X-rays (30, 31), which in turn modulates photoionization processes. In particular, this change in SED preferentially changes the relative strength between the ionizing (EUV/X-ray) continuum and optical/UV continuum, and therefore more greatly affects high-ionization lines than low-ionization lines. The idea of SED variations has been applied to explain the Baldwin effect of \CIV\ (62), and the same mechanism may apply to all emission lines that are powered by photoionization. In addition, there may be accompanying changes in the structure of the accretion flow and other physical processes such as accretion disk wind driving and X-ray gas shielding (33, 62, 63). One possibility is that the AD thickness as well as flaring towards the BH may change with Eddington ratio, which may lead to changes in the illumination configuration of the NLR. Another possibility is that changes in the accretion flow modify the structure of a possible optically thick gas component in the inner AD (64), which blocks some fraction of the emission from the inner accretion region and modifies the incident ionizing continuum as seen by the photoionized gas.

This simple framework can be extended to include BH mass as a third parameter to unify both high-luminosity quasars and low-mass broad-line AGNs with substantially lower luminosities than the quasar sample considered here. When such low-mass systems are included in the EV1 plane (Fig.\ 1), they will follow similar trends, with their [OIII] strength increased following the Baldwin effect. The nature of the Baldwin effect is unknown, but could be related to BH mass (or luminosity) that also plays a role in determining the accretion flow in addition to the Eddington ratio. 

Orientation plays an important role in broadening the distribution of FWHM$_\textrm{\hbeta}$, as expected from a flattened geometry for the BLR. The dispersion in FWHM$_\textrm{\hbeta}$ at fixed $\Rfe$ is largely dominated by the orientation effect; it does not reflect real changes in the underlying virial velocity of the BLR and therefore changes in the BH mass. Although the average FWHM$_\textrm{\hbeta}$ at fixed $\Rfe$ does correlate with the BH mass (as confirmed by clustering analysis), individual FWHM$_\textrm{\hbeta}$ measurements introduce substantial scatter in the virial BH mass estimates and dilute the intrinsic distinction in BH masses. The observed dependence of FWHM$_\textrm{\hbeta}$ on radio morphologies, the distribution of line shape in the EV1 plane, and the relative prominence of torus emission to disk emission at fixed $\Rfe$ are all consistent with this orientation interpretation. 

The postulated BLR disk should be coplanar with the AD (or the BLR itself could be an extension of the AD). Assuming the optical luminosity $L_{5100}$ comes from an optically-thick, geometrically-thin standard AD, the observed $L_{5100}$ should also have an orientation bias. This is hinted by the excess optical-IR color shown in Extended Data Fig.\ E6. In our scenario, at fixed $\Rfe$, the vertical bins in Fig.\ 1 represent a change in orientation of the BLR disk (and the AD). The change in the relative optical luminosity to IR luminosity along vertical bins is best explained by the orientation bias in optical AD luminosity (see Extended Data Fig.\ E6), assuming that the IR luminosity is isotropic. Denoting $I$ as the inclination angle between the normal of the disk and line-of-sight (LOS), the orientation bias reduces the AD luminosity by a factor of $\cos I$ from face-on. Since the reduction in the relative optical-IR luminosity is roughly 20-30\% ($\sim 0.2$ mag) across the vertical direction in Extended Data Fig.\ E6, this orientation bias in AD luminosity translates to $I_{\rm max}\approx 45^\circ$, where $I_{\rm max}$ is the maximum inclination for the quasar to be a broad-line object. This $I_{\rm max}$ corresponds to an average inclination angle of $\langle I\rangle=30^\circ$ (assuming random orientation of the disk normal to the LOS). All these are quite reasonable numbers for type 1 quasars. We also emphasize that because SDSS is a shallow flux-limited survey, most of the quasars in our sample are likely seen close to face-on, and the distribution of inclination may be quantitatively different from that of other samples with heterogeneous selection. In particular, an average inclination of $\langle I\rangle=30^\circ$ for the AD (and by extension, the postulated BLR disk) indicates a virial coefficient of $f_{\sigma}=8$ for the average conversion between LOS broad-line width (second moment, $\sigma$) and virial velocity $V$ (i.e., $\sigma^2=\frac{1}{2}V^2\sin^2 I=V^2/f_{\sigma}$, neglecting turbulent broadening). This virial coefficient is fully consistent with the empirically derived values of $f_\sigma\sim 6$ based on the BH mass-bulge scaling relations (50, 65), considering the fudge factors in the average inclination angle and the unknown turbulent broadening $s$ (e.g., in practice $\sigma^2=\frac{1}{2}V^2\sin^2I+s^2$, hence the actual virial coefficient $f_{\sigma}=V^2/\sigma^2<\frac{2}{\sin^2I}$). Therefore, the orientation bias from a thin standard AD only introduces $<30\%$ variation in $L_{5100}$ in the vertical bins in Fig.\ 1, which is consistent with no apparent trend in the [OIII] strength.

On the other hand, while the AD luminosity change due to orientation is small, the change in the line width will be substantial. The line width scales as $\sin I$. Hence from $I_{\rm max}=45^\circ$ to $\langle I\rangle=30^\circ$ there is a factor of $\sim 1.5$ change; changing from $\langle I\rangle$ to nearly face-on can yield an even larger reduction factor, but at very small inclinations, the isotropic velocity component (such as turbulence) in the BLR will start to dominate the line width. These estimates are consistent with the distribution in the 2D EV1 plane (Fig.\ 1). Although the exact values of $I_{\rm max}$ and $\langle I\rangle$ will be different, the approximate agreement between these simple estimates and observations is reassuring that our scenario is correct.

Higher-Eddington ratio quasars (with higher $\Rfe$) may drive stronger outflows in both the broad-line region and the narrow-line region. This is in line with the following observations: (a) the velocity offset of the wing component of \OIII\ is larger in quasars with higher $\Rfe$; (b) the fraction of quasars with broad absorption lines, which likely arise from a disk wind (66, 67), seems to increase at the high-$\Rfe$ end of EV1 (4), and broad absorption-line quasars usually have weak \OIII\ lines; (c) the enhancement of torus emission relative to AD emission at the high-$\Rfe$ end of EV1 may be caused by more efficient disk winds that facilitate the formation of a dusty torus (68). In addition, we point out that FWHM$_\textrm{\hbeta}<2000\,{\rm km\,s^{-1}}$ should not be the sole criterion to define a distinct narrow-line Seyfert 1 population, as done in some studies. If the narrow-line Seyfert 1 phenomenon (69) is associated with one extreme end of EV1 (4), then the classification criteria must also include strong optical \FeII\ lines as well as weak \OIII\ emission.

Finally we comment on the implication of our framework on the frequency of quasars with double-peaked broad-line profiles, dubbed ``disk emitters'', which are characteristic of line emission from a Keplerian disk (70). The frequency of disk emitters is much higher in radio-loud quasars (71) ($\sim 10-15\%$) than in radio-quiet quasars (7, 72) ($\sim 3\%$). Our framework suggests a flattened geometry for the BLR in general, but only compact BLRs will show an apparent double-peaked profile in disk models (70, 73), suggesting that the BLR in the majority of quasars has a large extent in disk radii. Radio-loud quasars are preferentially low-Eddington ratio systems compared to the general quasar population (45), and hence their BLR size on average should be relatively smaller (in units of the gravitational radius of the BH) given the observed empirical relation between BLR size and luminosity (27). This may explain why radio-loud quasars are more often to show double-peaked broad-line profiles indicative of compact BLR disks.


\end{document}